\documentclass[12pt]{elsarticle}                                          
\usepackage{graphicx}                                                         
\usepackage{a41}                                                                
\usepackage{xcolor}                     
\usepackage[rflt]{floatflt}
\usepackage{float}
\usepackage{lscape}
\usepackage{array}
\usepackage[english]{babel}
\usepackage[T1]{fontenc}
\usepackage{ae}
\usepackage{url}
\usepackage{amsmath, amsthm, amssymb}
\usepackage{slashed}

\usepackage{rotating}
\usepackage{graphicx}
\usepackage{comment}
\newcounter{mmacnt}
\def\restartmma{\setcounter{mmacnt}{0}}
\restartmma \catcode`|=\active
\def|#1|{\mathrm{#1}}
\catcode`|=12
\newenvironment{mma}{
\par\smallskip
\catcode`|=\active
\parskip=0pt\parindent=0pt % locally
\small
\def\In##1\\{%
\def\linebreak{\hfill\break\null\qquad}%
\refstepcounter{mmacnt}
\hangindent=2.5em\hangafter=0
\leavevmode
\llap{\tiny\sffamily In[\arabic{mmacnt}]:=\kern.5em}%
\mathversion{bold}\footnotesize$
\displaystyle##1$\normalsize
\mathversion{normal}\par
 }%
\def\Print##1\\{%
\def\linebreak{\hfill\break}%
\hangindent=2.5em\hangafter=0
\leavevmode ##1\par}%
\def\Out##1\\{%
\def\linebreak{$\hfill\break\null\hfill$}%
\kern\abovedisplayskip\par
\hangindent=2.5em\hangafter=0
\leavevmode
\llap{\tiny\sffamily Out[\arabic{mmacnt}]=\kern.5em}
\footnotesize$\displaystyle##1$
\normalsize\hfill\null\par
\kern\belowdisplayskip
}%
\def\Warning##1##2\\{%
\def\linebreak{\hfill\break}%
\hangindent=2.5em\hangafter=0
\leavevmode
{\scriptsize##1 : ##2}\par}%
}{%
\par\smallskip
}

\usepackage{color}
\newenvironment{fshaded}{%
\MakeFramed {\FrameRestore}
}%
{\endMakeFramed}

\makeatletter
\def\ps@pprintTitle{%
\let\@oddhead\@empty
\let\@evenhead\@empty
\def\@oddfoot{\reset@font\hfil\thepage\hfil}
\let\@evenfoot\@oddfoot
}
\makeatother
\usepackage{tikz}
\usetikzlibrary{matrix}
\allowdisplaybreaks[4]

\begin{document}
%%%%%%%%%%%%%%%%%%%%%%
\begin{frontmatter}%%%
%%%%%%%%%%%%%%%%%%%%%%
\title{\Large
\textbf{
Production of charged Higgs bosons
associated with CP-even Higgs bosons
at future multi–TeV muon colliders
}}
%%%%%%%%%%%%%%%%%%%%%%%%%%%%%%%%%%%%%%%%%%%%%%
\author[1]{Quang Hoang-Minh Pham}
\author[1]{Khoa Ngo-Thanh Ho}
\author[2,3]{Khiem Hong Phan}
\ead{phanhongkhiem@duytan.edu.vn}
%%%%%%%%%%%%%%%%%%%%%%%%%%%%%%%%%%%%%%%%%%%%%%
\address[1]
{\it VNUHCM-University of Science,
$227$ Nguyen Van Cu, District $5$,
Ho Chi Minh City $700000$, Vietnam}
\address[2]{\it Institute of Fundamental
and Applied Sciences, Duy Tan University,
Ho Chi Minh City $70000$, Vietnam}
\address[3]{Faculty of Natural Sciences,
Duy Tan University, Da Nang City $50000$,
Vietnam}
%%%%%%%%%%%%%%%%
\pagestyle{myheadings}
\markright{}
%%%%%%%%%%%%%%%%%%%%
\begin{abstract} %%%
%%%%%%%%%%%%%%%%%%%%
We report the first results for charged Higgs boson
production associated with CP-even Higgses at future
multi–TeV muon colliders within the scenario of the
Type-Y Two-Higgs-Doublet Model (THDM). The valid
parameter regions in the Type-Y THDM are first
updated, and based on the allowed parameter space,
all two-body decay channels of the charged Higgs
boson are then computed with the help of {\tt
H-COUP}. Both production processes $\mu^- \mu^+ \to
H^\pm H^\mp h, H^\pm H^\mp H \to t\bar{t} b\bar{b} h, 
t\bar{t} b\bar{b}H$ are
scanned over the constrained parameter space of the
considered model. The production cross sections can
reach $\sim 0.5$ fb in several regions of the
parameter space. The results provide an opportunity
to probe charged Higgs bosons produced in association
with the heavy CP-even Higgses at the high integrated
luminosities proposed for future multi–TeV muon
colliders. Furthermore, we evaluate the signal
significances for $\mu^- \mu^+ \to H^\pm H^\mp h \to
t\bar{t} b\bar{b} h \to t\bar{t} b\bar{b} b\bar{b}$
with respect to the Standard Model backgrounds. We
also estimate the significances by including the top
quark decay into leptons and bottom quarks and 
considering $b$-tagging. Thanks to the high integrated
luminosities anticipated at future multi–TeV muon
colliders,we find that the signal significances 
can exceed $\sim 3\sigma$ at several benchmark points
selected in the viable parameter space of the Type-Y THDM.
%%%%%%%%%%%%%%%%%%%%
\end{abstract}
%%%%%%%%%%%%%%%%%
\end{frontmatter}
%%%%%%%%%%%%%%%%%%%%%%%%%%%
\section{Introduction}%%%%%
%%%%%%%%%%%%%%%%%%%%%%%%%%%
The scalar Higgs sector remains one of the least understood parts
of the Standard Model. It is known that experimental programs 
at future colliders,
such as the High-Luminosity Large Hadron Collider (HL-LHC), the
High-Energy LHC (HE-LHC), as well as proposed future lepton
colliders, including the International Linear Collider and
multi--TeV muon colliders, are planned to explore the structure
of the scalar Higgs sector. In pursuing the goals, the program 
involves not only precise measurements of the SM-like
Higgs boson properties but also searches for additional scalar
particles at future colliders. In many of the additional 
scalar production processes, searches for singly charged 
Higgs bosons are interest greatly at current and future colliders. 
We briefly summarize several representative works in this section. 
At $pp$ collisions, the light charged Higgs mass region has been 
measured produced via top-quark decay channels, 
as reported in Refs.~\cite{CMS:2012fgz,ATLAS:2023bzb,
ATLAS:2024oqu,CMS:2015lsf}.
Charged Higgs bosons associated with top--pair production 
have been investigated at $\sqrt{s}=7~\text{TeV}$ at the LHC, 
including the sequential decay 
mode $H^\pm \to \tau \nu_{\tau}$ in Refs.~\cite{ATLAS:2012nhc,ATLAS:2018gfm,
ATLAS:2024hya,CMS:2019bfg}, and the decay channel $H^\pm \to c\bar{s}$ in Refs.~\cite{ATLAS:2013uxj,CMS:2020osd}.
Searches for a heavy charged Higgs mass regions 
via the decay $H^\pm \to tb$ through the scattering process 
$pp \to tH^\pm \to t\bar{t}b$ have also been reported 
at the LHC~\cite{ATLAS:2015nkq,ATLAS:2021upq,CMS:2020imj}.
Furthermore, probing charged Higgs production 
via vector boson fusion,
including the decay $H^\pm \to W^\pm Z$ as
published in Ref.~\cite{ATLAS:2015edr, CMS:2021wlt}.
A rare decay mode of the charged Higgs into charm and bottom quarks
in Ref.~\cite{CMS:2018dzl} and the decay channel $W^\pm h$
in \cite{CMS:2022jqc, ATLAS:2024rcu} have been searched 
in proton–proton collisions.
%%%%%%%%%%%%%%%%%%%%%%%%%%%%%%%%%%%%%%%%%%%%%%%%%%%%%%%%%%%%%%%%%

Many theoretical computations for charged Higgs productions 
at the LHC within various extensions of the SM have been 
performed. In Refs.~\cite{Arhrib:2017veb,
Arhrib:2018bxc}, the charged Higgs productions
together with a top quark and its decay into a top and bottom quark,
including top polarization effects, have been studied at the LHC.
Moreover, charged Higgs boson pair production including 
its bosonic decay channels at the HL-LHC as well as the HE-LHC 
have been reported in~\cite{Arhrib:2019ywg}. 
The computation of charged Higgs production within the
Minimal Supersymmetric Standard Model 
has been considered in Ref.~\cite{Arhrib:2019ykh}.
Furthermore, the implications for charged Higgs at 
the LHC in the different mass regions have been investigated as
in Refs.~\cite{Arhrib:2020tqk,Arhrib:2021xmc, 
Wang:2021pxc,Krab:2022lih, Arhrib:2024sfg, Arhrib:2024nbj,
Logan:2018wtm}. Charged Higgs productions 
have also been studied at future lepton colliders (LCs), 
including $e^-e^+$ collisions and
multi--TeV muon colliders, as reported in 
Refs.~\cite{Ouazghour:2023plc,Ouazghour:2024twx,
Ouazghour:2025owf, BrahimAit-Ouazghour:2025mhy,
Ahmed:2024oxg, Hashemi:2023osd}
and also in many previous
papers~\cite{Tran:2025iur,Tran:2025zfq,
Phan:2025pjt}. Future LCs are 
proposed for high-precision tests and searches for additional 
scalar particles in many beyond the Standard Models (BSM). 
The LCs provide a cleaner environment than the LHC, which 
suffers from huge QCD backgrounds. Moreover, 
multi--TeV muon colliders are designed for
higher frontier energy regimes for testing new physics. In
comparison with $e^- e^+$ collisions, a key advantage of
multi--TeV muon colliders is that the muon mass is about 207
times larger than the electron mass. Subsequently, s-channel
exchanges by new scalar particles may exist and can be enhanced
due to resonance effects. Furthermore, the couplings of the
scalar to muons are proportional to the mixing angle $\beta$.
Depending on the type of THDM, these contributions can also be
significant. For the above reasons, multi--TeV muon colliders
offer a great opportunity to probe and distinguish different
types of THDM. In many of our previous 
works~\cite{Tran:2025iur,Tran:2025zfq,Phan:2025pjt}, we have
studied charged Higgs boson production in Type-I and Type-X
THDM. In this work, we report the first results for charged
Higgs boson production associated with CP-even Higgses 
at future multi--TeV muon colliders in the scenario of
the Type-Y THDM. The valid parameter regions for
the Type-Y THDM are updated in this study. Based on 
the allowed parameter space, we compute all two-body 
decay channels of the charged Higgs boson
with the help of {\tt H-COUP}~\cite{Aiko:2023xui,Aiko:2021can}. 
The production cross sections
for charged Higgs boson production associated with CP-even Higgs
states are scanned over the viable parameter space at $3$ TeV
and $5$ TeV center-of-mass energies at future multi--TeV
muon colliders. Furthermore, we compute the signal significances
for $\mu^- \mu^+ \to H^\pm H^\mp h \to t b t b h$ in the
sequential decay of ${h \to b\bar{b}}$ with respect to the
Standard Model background. We
also estimate the significances by including the top
quark decay into leptons and bottom quarks and 
considering $b$-tagging. Thanks to the high integrated
luminosities anticipated at future multi–TeV muon
colliders,we find that the signal significances 
can exceed $\sim 3\sigma$ at several benchmark points
selected in the viable parameter space of the Type-Y THDM.
%%%%%%%%%%%%%%%%%%%%%%%%%%%%%%%%%%%%%%%%%%%%%%%%%%%%%%%%%%%%

The remaining sections of the paper are organized as follows.
In Section~2, the THDM  are briefly reviewed and 
its constraints are discussed in further detail. 
Section~3 presents detailed calculations and
phenomenological results for $\mu^- \mu^+ \to H^\pm H^\mp h/H$
at multi--TeV muon colliders. Conclusions and outlook are 
presented in Section~4. The Appendix shows 
the amplitudes for the processes under investigation.
%%%%%%%%%%%%%%%%%%%%%%%%%%%%%%%%%%%%%%%
\section{The Two-Higgs-Doublet Model
and Its Constraints}%%%%%%%%%%%%%%%%%%%
%%%%%%%%%%%%%%%%%%%%%%%%%%%%%%%%%%%%%%%
In this section, we briefly review the THDM.
We refer to our previous works~\cite{Tran:2025iur,Tran:2025zfq,
Phan:2025pjt} and Ref.~\cite{Branco:2011iw} for a comprehensive 
review of the THDM. In the studied model, the matter and gauge 
boson content remain the same as in the SM, while the scalar 
sector is extended by an additional scalar doublet with 
hypercharge $Y = 1/2$. Consequently, the scalar potential and 
Yukawa Lagrangian are enlarged compared with those of the SM. 
The scalar potential is given as follows in Ref.~\cite{Tran:2025iur}:
%%%%%%%%%%%%%%%%%%%%%%%%%%%%%%%%%%%
\begin{eqnarray} 
\label{potential}
\mathcal{V}
(\Phi_1, \Phi_2) &=&
\sum\limits_{j=1}^2
m^2_{jj}\Phi_j^\dagger\Phi_j
-
\left(m^2_{12}\Phi_1^\dagger\Phi_2
+
{\rm H.c.}\right)
+
\frac{1}{2}
\sum\limits_{j=1}^2
\lambda_j\left(\Phi_j^\dagger\Phi_j\right)^2
\nonumber \\
&&
\hspace{0cm}
+\lambda_3\Phi_1^\dagger\Phi_1\Phi_2^\dagger\Phi_2
+\lambda_4\Phi_1^\dagger\Phi_2\Phi_2^\dagger\Phi_1
+\left[\frac{1}{2}
\lambda_5\left(\Phi_1^\dagger\Phi_2\right)^2
+{\rm H.c.}\right].
\end{eqnarray}
%%%%%%%%%%%%%%%%%%%%%%%%%%%%%%%%%%%
As in Ref.~\cite{Phan:2025pjt}, we also focus on the CP-conserving version of the THDM, in which all parameters, such as $m_{11}^2$, $m_{22}^2$, $m_{12}^2$, $\lambda_1$, $\ldots$, $\lambda_5$, are taken to be real variables. As indicated in many previous works, a $Z_2$ discrete symmetry is introduced into the scalar potential (up to the soft-breaking term) to avoid tree-level flavor-changing neutral currents. Consequently, the THDM is classified into four distinct types, which have different Yukawa couplings as shown in Ref.~\cite{Aoki:2009ha}.
The Yukawa Lagrangian is written in its most general 
parametrized form, as presented in Ref.~\cite{Tran:2025iur}:
%%%%%%%%%%%%%%%%%%%%%%%%%%%%%%%
\begin{eqnarray}
\label{Yukawa}
-{\mathcal L}_\text{Y} 
&=&
\sum_{f=u,d,\ell}
\left(
\sum_{\phi_j=h, H}
\frac{m_f}{v}\xi_{\phi_j}^f
\phi_j {\overline f}f
-i\frac{m_f}{v}\xi_A^f
{\overline f}
\gamma_5fA
\right)
\\
&&
+
\frac{
\sqrt{2}
}{v}
\left[
\bar{u}_{i}
V_{ij}\left(
m_{u_i}
\xi^{u}_A P_L
+
\xi^{d}_A
m_{d_j} P_R \right)d_{j} H^+
\right]
% \nonumber\\
% &&
+ \frac{\sqrt{2}}{v}
\bar{\nu}_L
\xi^{\ell}_A
m_\ell \ell_R H^+
+ \textrm{H.c}.
\nonumber
\end{eqnarray}
%%%%%%%%%%%%%%%%%%%%%%%%%%%%%%%%%
The elements of the CKM matrix are denoted 
by $V_{ij}$, and the left- and right-handed leptons 
are represented as $\ell_{L/R}$. The projection 
operators are defined as
$P_{L/R} = \frac{1 \mp \gamma_5}{2}$.
Allcoupling coefficients appearing in 
Eq.~\ref{Yukawa} are summarized in 
Table~2 of Ref.~\cite{Aoki:2009ha}.
%%%%%%%%%%%%%%%%%%%%%%%%%%%%%%%%%%%
After electroweak symmetry breaking (EWSB), the additional scalar particles in the considered model include a CP-even Higgs ($H$),
a CP-odd Higgs ($A$), and two singly charged Higgs bosons ($H^\pm$). The independent parameters of the THDM used in our analysis consist of the mixing angles, expressed in terms of $s_{\beta-\alpha}$ and $t_\beta$, the scalar masses $m_H$, $m_A$, $m_{H^\pm}$, and the soft-breaking parameter $m_{12}^2$. 

%%%%%%%%%%%%%%%%%%%%%%%%%%%%%%%%%%%%%%%%%%%%%%%%%%%%%%%
Before presenting the calculations and 
phenomenological results of this work, we
first update the parameter constraints for the Type--Y THDM in the
following paragraphs. Detailed explanations of the theoretical and
experimental constraints were given in our previous studies, especially
for Type--I in Ref.~\cite{Tran:2025iur} and for 
Type--X in Ref.~\cite{Tran:2025zfq,
Phan:2025pjt}. The present
work focuses on scanning the parameter space of the Type--Y THDM. The
parameter space of the Type--Y THDM is scanned over the same ranges
as in Refs.~\cite{Tran:2025iur}: $s_{\beta-\alpha} \in [0.97, 1]$, $t_{\beta} \in
[0.5, 45]$, $m_{H} \in [130, 1000]~\text{GeV}$, $m_{A,H^{\pm}} \in
[130, 1000]~\text{GeV}$, and $m_{12}^2 \in [0, 10^6]~\text{GeV}^2$, with
the SM-like Higgs mass fixed at $m_{h} = 125.09~\text{GeV}$.
It is emphasized that the scalar masses in the Type-Y THDM should be scanned up to 1500 GeV. However, in this study, we focus on charged Higgs bosons with masses up to 1000 GeV. Heavier charged Higgs states generally yield smaller production cross sections, making them difficult to probe at future colliders. For these reasons, we restrict our parameter space to the mass ranges specified above.
The steps for imposing the parameter constraints can be summarized as
follows. First, the sampled points are checked against the theoretical
conditions. In this step, the model under consideration must satisfy
unitarity, perturbativity, and vacuum stability of the scalar potential.
Second, the surviving parameter points are tested against the experimental
limits on the $S$, $T$, and $U$ parameters at the $95\%$~CL. In the next
step, all allowed points are confronted with the LHC data. We note that the computer program {\tt 2HDMC}-1.8.0~\cite{Eriksson:2009ws} is used for all the above tests. In the third stage of constraints, both {\tt HiggsSignals}
~\cite{Bechtle:2013xfa}
and {\tt HiggsBounds}~\cite{Bechtle:2020pkv}
are taken into account for matching the theoretical
predictions with the corresponding experimental data. 
Their implementations 
are already included in {\tt 2HDMC}-1.8.0. Finally, 
all the remaining points are 
further constrained by the flavor data, 
which can be evaluated using 
{\tt SuperISO}~\cite{Mahmoudi:2008tp}.

\begin{figure}[]
\centering
\begin{tabular}{cc}
\includegraphics[width=8cm, height=5cm]
{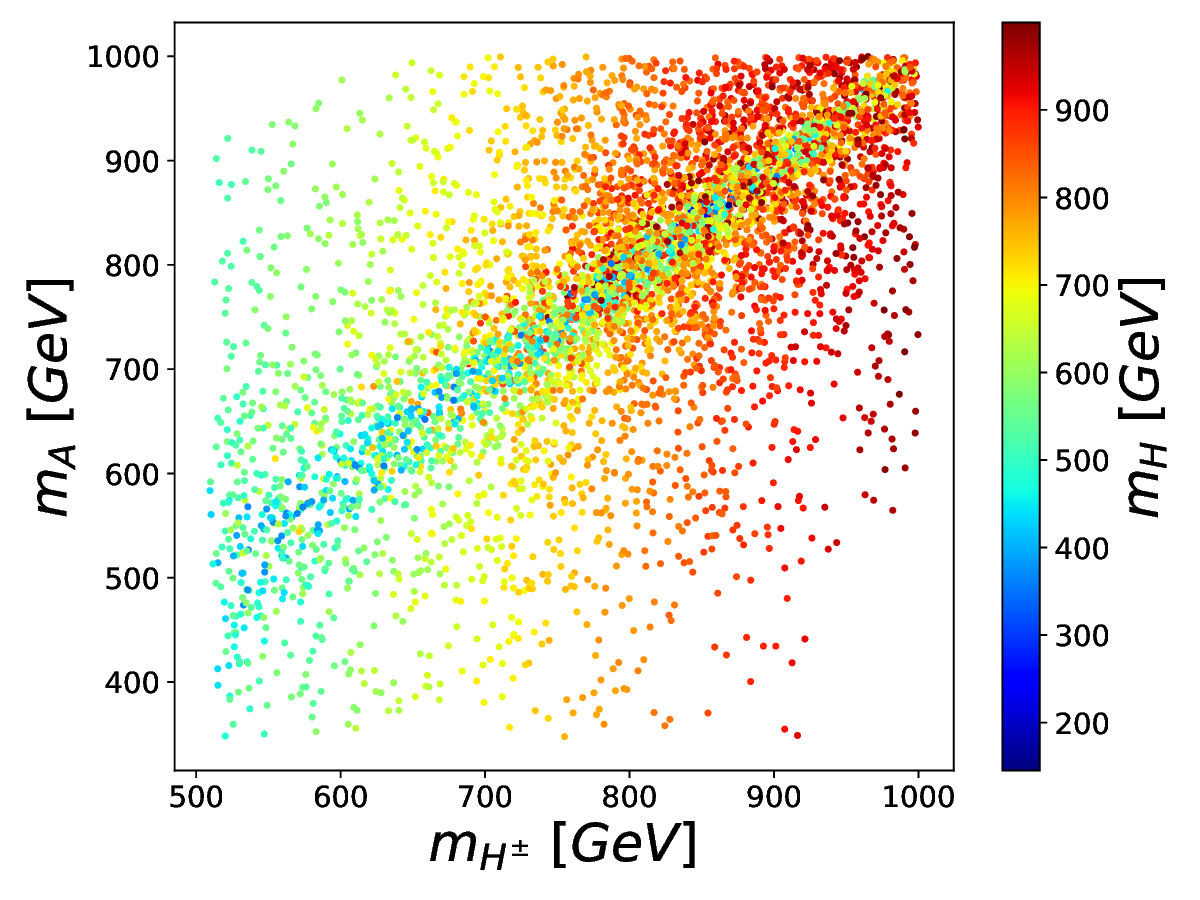}
&
\includegraphics[width=8cm, height=5cm]
{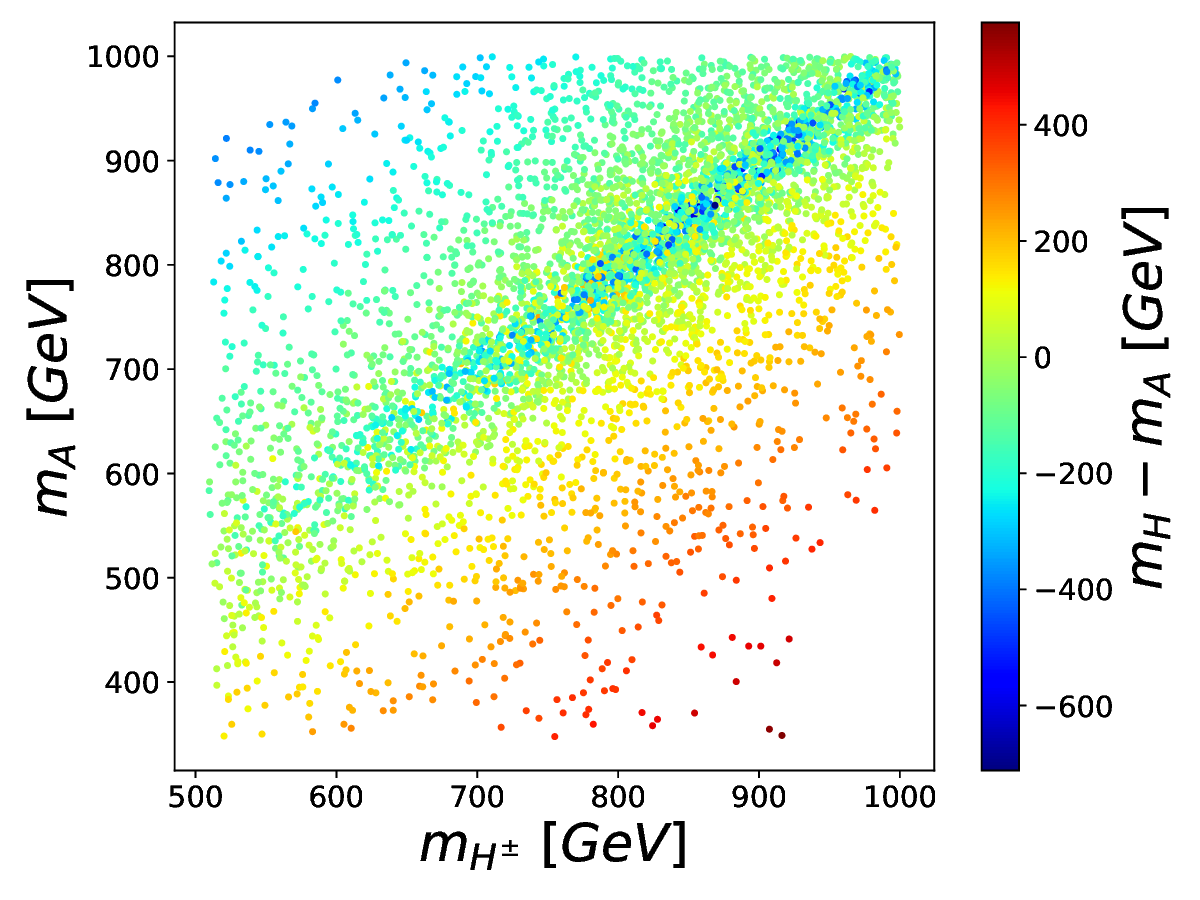}
\\
\includegraphics[width=8cm, height=5cm]
{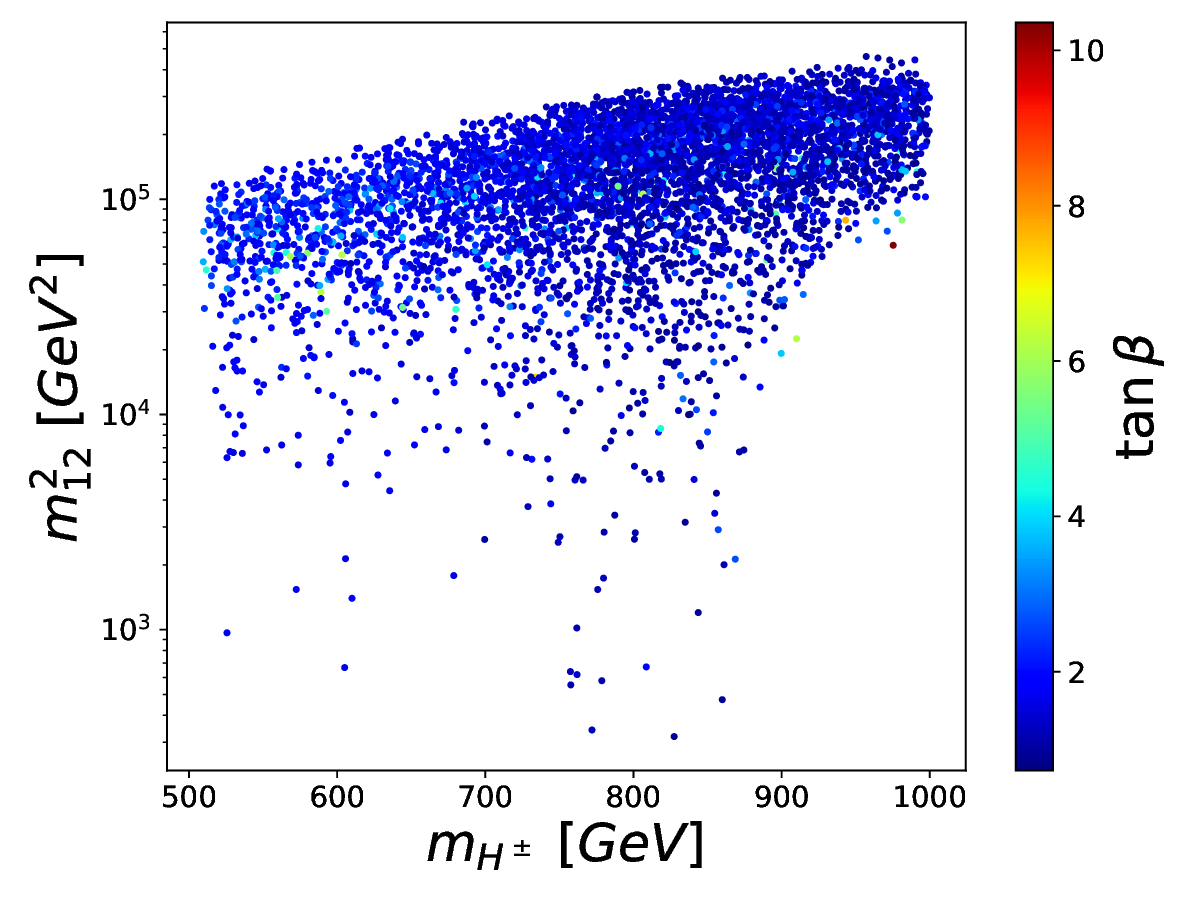}
&
\includegraphics[width=8cm, height=5cm]
{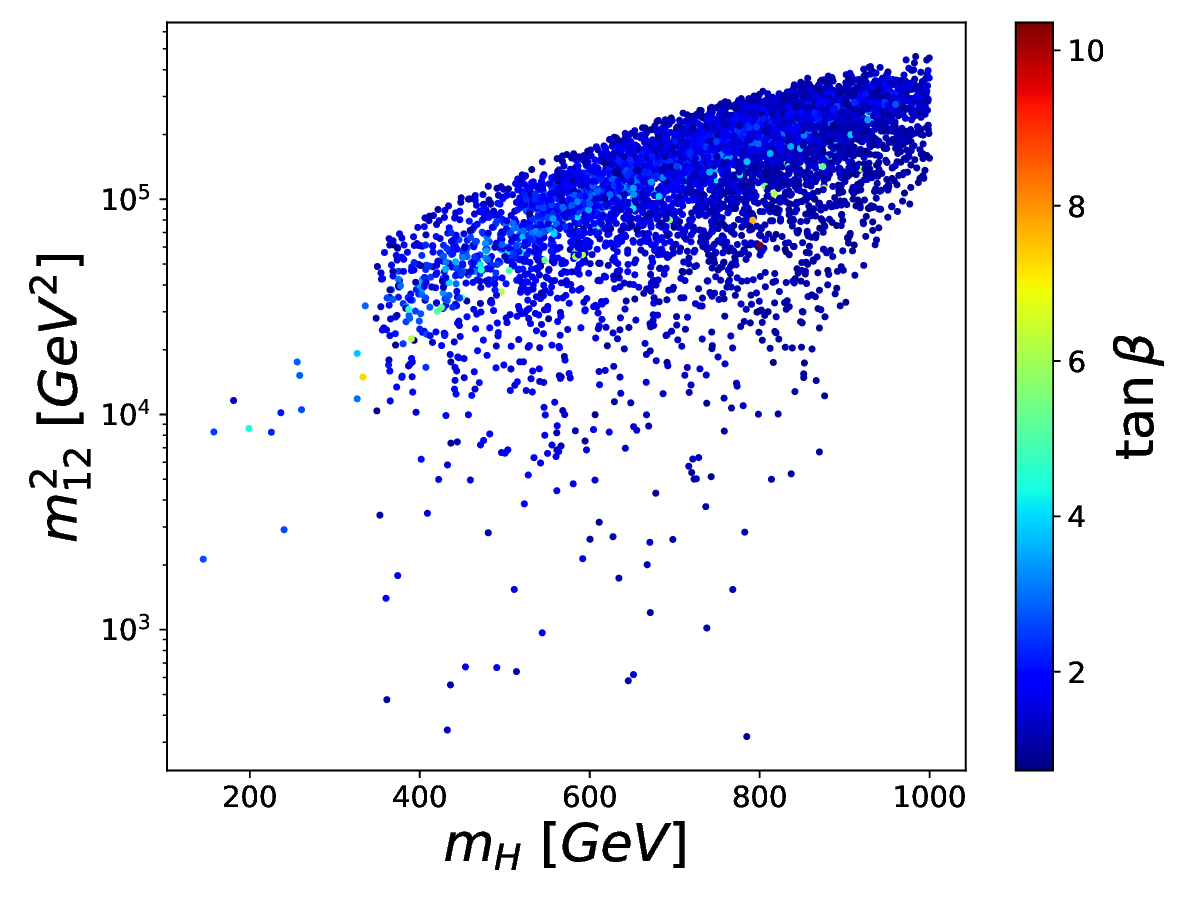}
%%%%%%%%%%%%%%%%%%%%%%%%%%%%%%%%%%%%%%%
\end{tabular}
\caption{\label{scan1}
The scatter plots show the correlations among $m_A$,
$m_{H^\pm}$, and $m_H$ in the upper-left panel, and
those among $m_A$, $m_{H^\pm}$, and $m_H - m_A$ in
the upper-right panel. The lower-left plot shows the
correlations between $m_{12}^2$, $m_{H^\pm}$, and
$\tan\beta$, while the lower-right panel illustrates
the correlations among $m_{12}^2$, $m_H$, and
$\tan\beta$.}
\end{figure}

After obtaining the valid parameter space for
the Type--Y THDM, we proceed to discuss the physical
regions through the following scatter plots. We first
focus on the scatter plot showing the correlations
among $m_A$, $m_{H^\pm}$, and $m_H$ in the upper-left
panel, and those among $m_A$, $m_{H^\pm}$, and
$m_H - m_A$ in the upper-right panel, as shown in
Fig.~\ref{scan1}. From these plots, we observe that
in the region where $m_A \simeq m_{H^\pm}$, the CP-even 
Higgs $H$ can span a wide range from approximately 
$200~\text{GeV}$ to $1000~\text{GeV}$. The second 
favored regime corresponds to $m_H \simeq m_{H^\pm}$, 
in which the CP-odd Higgs $A$ is allowed to take
any value from roughly $200~\text{GeV}$ to $1000~\text{GeV}$.
Flavor constraints require the charged Higgs mass to be
larger than $\sim 520$ GeV. In the upper-right panel plot, 
we find that for $m_A - m_H \sim -200~\text{GeV}$, the CP-odd
Higgs $A$ has a mass in the range of approximately 
$800~\text{GeV}$ to $1000~\text{GeV}$, while
$520~\text{GeV} \leq m_{H^\pm} \leq 700~\text{GeV}$.
Following the data shown in the left plot, if $m_A 
\simeq m_{H^\pm}$, the degenerate scalar mass difference
$m_A - m_H$ can span a wide range. Finally, one finds
that in the region $m_{H^\pm} \leq m_A$, the degenerate 
scalar mass difference ranges from $0~\text{GeV}$ to 
approximately $+200~\text{GeV}$.
The scatter plots show
the correlations between $m_{12}^2$, $m_{H^\pm}$, and
$\tan\beta$ in the lower-left panel, while the
correlations among $m_{12}^2$, $m_H$, and $\tan\beta$
are illustrated in the lower-right panel. Our finding
is that the updated parameter space is concentrated in
the region $10^4 \lesssim m_{12}^2 \lesssim 5 \cdot
10^{5}$ and $m_H \gtrsim 400~\text{GeV}$.
The favored data points concentrate in the 
region $t_{\beta} \leq 3$. Based on the resulting 
physical parameter regions,
we proceed to study the production of charged Higgs 
bosons in association with CP-even Higgs bosons at future
multi–TeV muon colliders in the following sections.
%%%%%%%%%%%%%%%%%%%%%%%%%%%%%%%%%%%%%%%
% %%%%%%%%%%%%%%%%%%%%%%%%%%%%%%%%%%%%%%%%%%%%%
\section{Production processes
$\mu^-\mu^+ \to H^\pm H^\mp h/H$ 
at multi--TeV muon colliders}
%%%%%%%%%%%%%%%%%%%%%%%%%%%%%%%%%%%%%%%%%%%%%%%
Detailed calculations for the production processes 
$\mu^-\mu^+ \to H^\pm H^\mp h/H$ at multi--TeV muon 
colliders are presented in this section. 
The calculations are performed with the help of 
the programs {\tt FeynArts/FormCalc}~\cite{Hahn:2000kx}. 
As stated in the introduction, the $s$-channel 
exchange of scalar particles may give significant 
contributions due to the resonance effects. 
This highlights the advantage of future multi--TeV 
muon colliders in comparison with $e^- e^+$ colliders.
In order to taking into account these contributions, 
we should consider all tree-level Feynman diagrams 
contributing to the production cross sections. 
It should be mentioned that the processes are generated
in the general $\mathcal{R}_{\xi}$ gauge.
The advantage of this choice is that the results
can be verified by self-consistency checks 
through their independence from the $\xi$-gauge 
parameter. 

Before scanning the production cross sections
over the valid parameter space of Type-Y THDM,
we first present the evaluations of the charged Higgs
branching fractions. The results from charged 
Higgs branching ratios allow us to select
the appropriate decay modes for the significance
simulations in the last subsection.
%%%%%%%%%%%%%%%%%%%%%%%%%%%%%%%%%%
\subsection{Branching fractions}%%
\begin{figure}[]
\centering
\begin{tabular}{cc}
\includegraphics[width=8cm, height=5cm]
{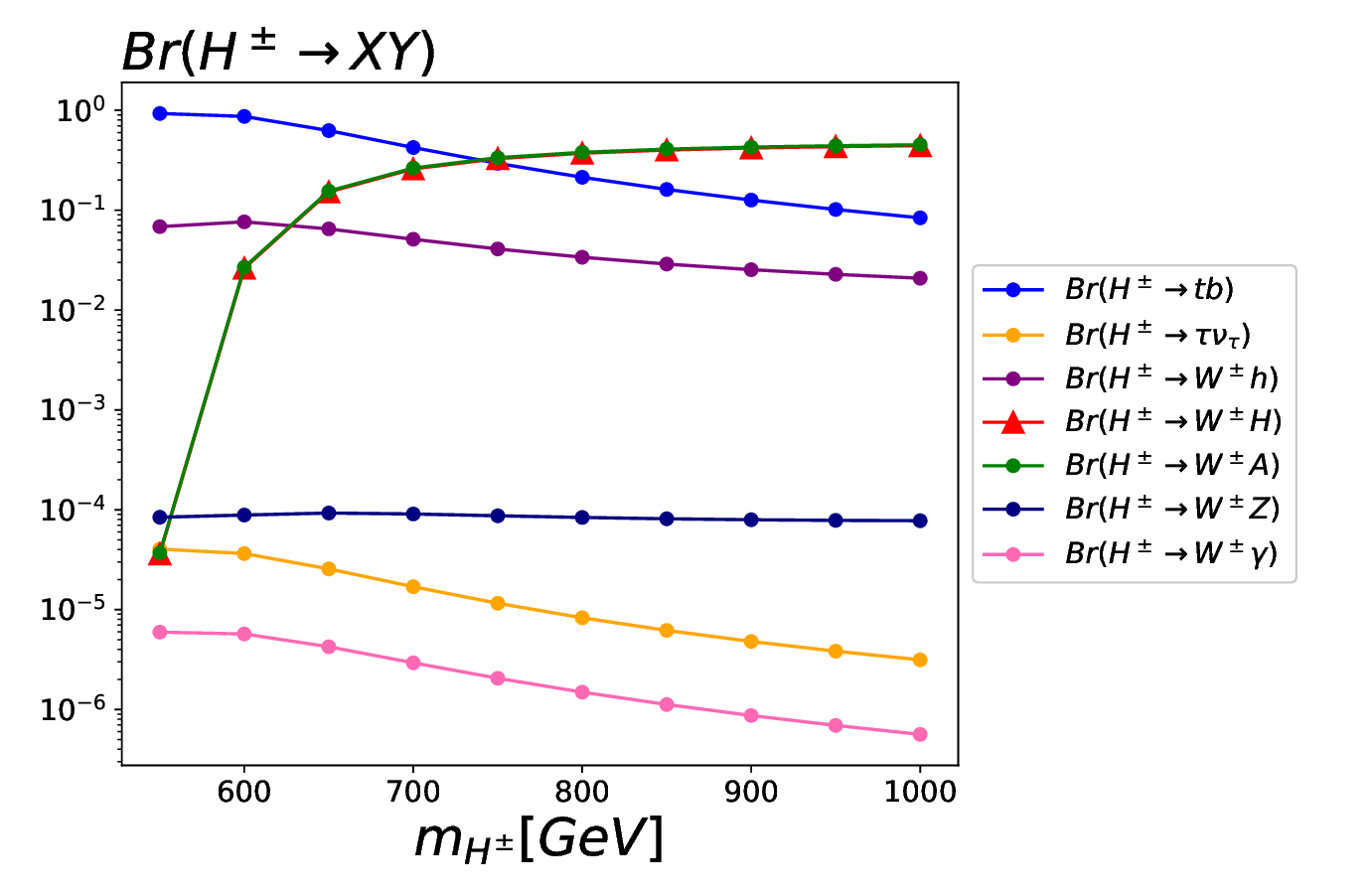}
&
\includegraphics[width=8cm, height=5cm]
{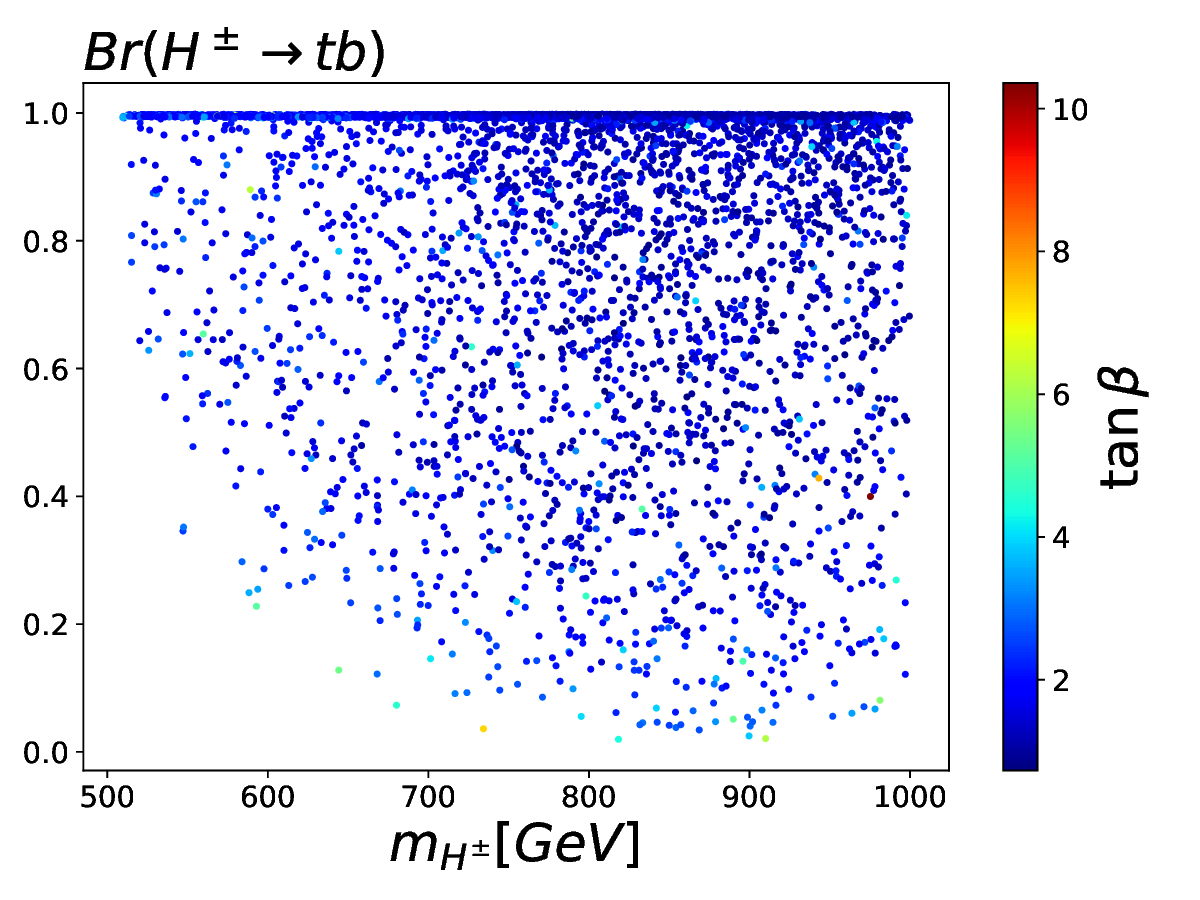}\\
\includegraphics[width=8cm, height=5cm]
{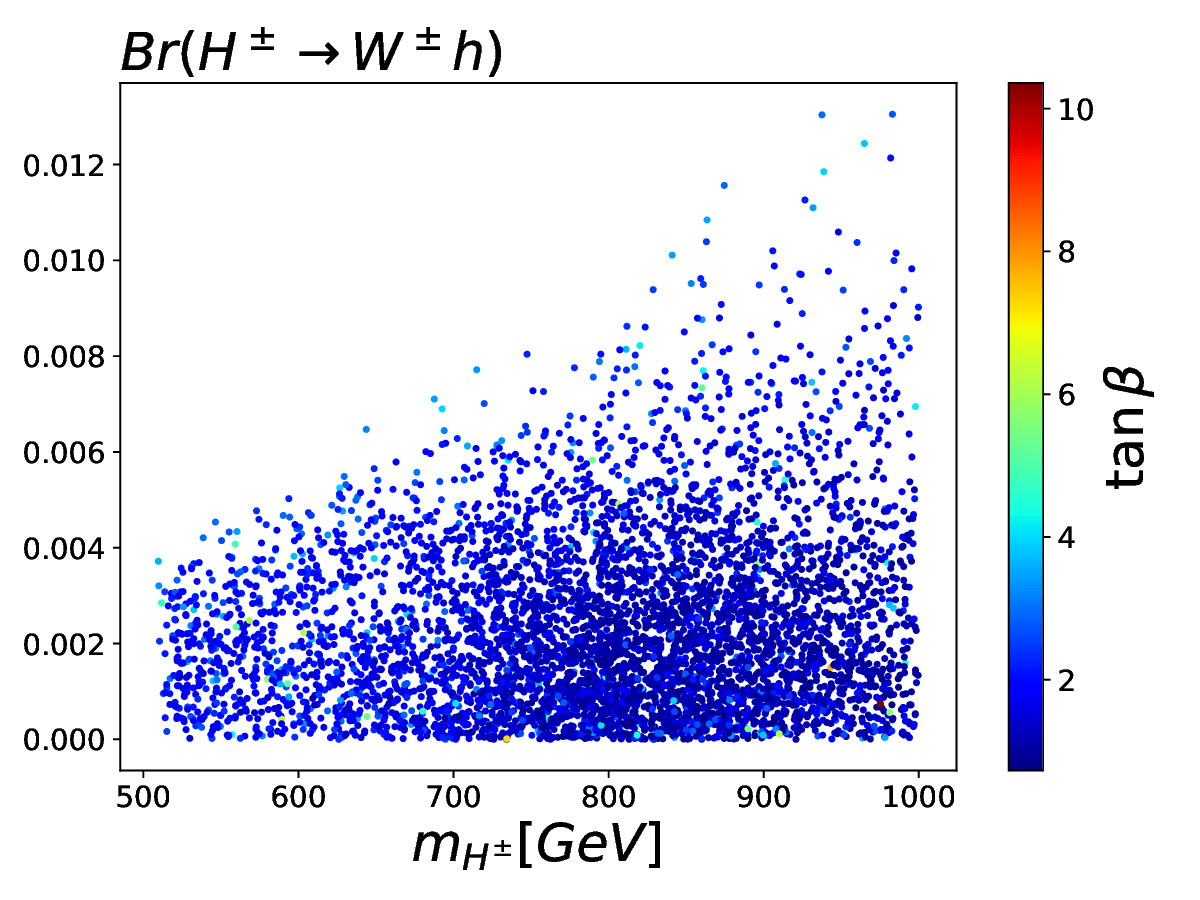}
&
\includegraphics[width=8cm, height=5cm]
{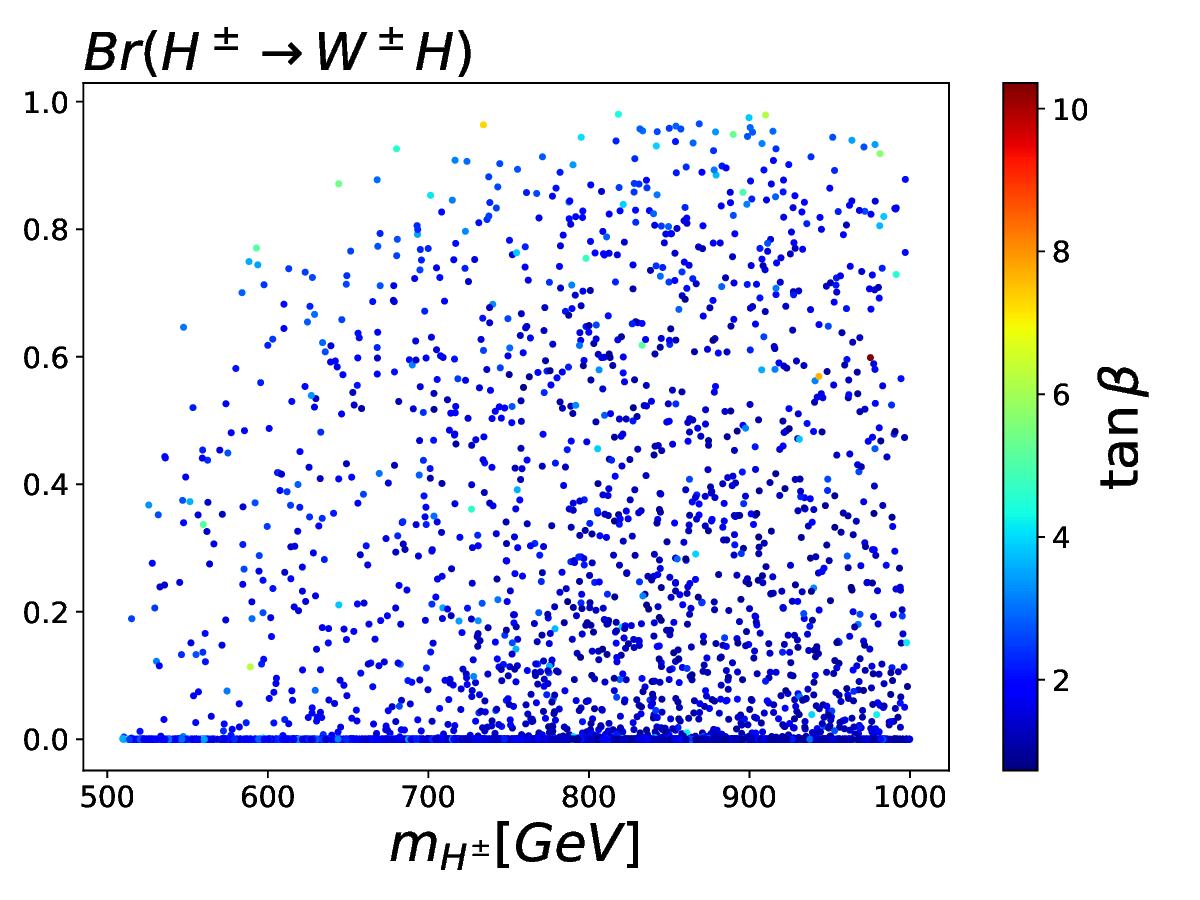}
%%%%%%%%%%%%%%%%%%%%%%%%%%%%%%%%%%%%%%%
\end{tabular}
\caption{\label{Br1}
The upper-left plot displays the branching 
fractions of the charged Higgs 
for all decay channels
at the benchmark frist point presented in Table~\ref{BP}.
The upper-right plot shows the decay rate of $H^\pm \to tb$, 
while the lower-left plot corresponds to $H^\pm \to W^\pm h$.
The lower-right plot illustrates $H^\pm \to W^\pm H$.
All plots are generated as functions of the 
charged Higgs mass and $t_{\beta}$.
}
\end{figure}

%%%%%%%%%%%%%%%%%%%%%%%%%%%%%%%%%%
The viable parameter space after the constraints
is transfered to {\tt H-COUP}~\cite{Aiko:2023xui,Aiko:2021can} 
for evaluating all two-body decay channels of 
charged Higgs. Its branching fractions for 
all the decay channels are shown in the 
following paragraphs. 
First, the branching ratios are 
geneated as functions of the charged Higgs 
mass at the benchmark first point
(BP1) showing in Table~\ref{BP}
for an example, as indicated in the above-left plot 
of Fig.~\ref{Br1}. Across the entire range of 
$m_{H^\pm}$, the decay $H^\pm \to tb$ 
dominates in the lower mass region of 
the charged Higgs below $\sim 750$ GeV. 
In the higher mass region ($m_{H^\pm}\geq \sim 
750$ GeV), the channels $H^\pm \to W^\pm \phi_j$ 
with $\phi_j = h, H, A$ open and they become 
the dominant contributions. Because 
the tree-level couplings of $H^\pm W^\mp H$ 
and $H^\pm W^\mp A$ differ only by the 
factor $s_{\beta-\alpha}$, and in this 
benchmark point we take $s_{\beta-\alpha} = 0.99$,
the branching ratios of $H^\pm \to W^\pm H$ 
and $H^\pm \to W^\pm A$ become nearly equal.
Furthermore, we find that the branching 
ratio of $H^\pm \to W^\pm h$ is approximately
$10^{-1}$ over the entire range of charged 
Higgs masses. In the above-right plot
of Fig.~\ref{Br1}, 
the branching fraction of $H^\pm \to t b$ 
is shown as a function of $m_{H^\pm}$ and 
$\tan\beta$. This decay mode dominates at low 
values of $\tan\beta$ and in the low-mass region of
the charged Higgs. As pointed out above,
the channels $H^\pm \to W^\pm h/H/A$ provide
large contributions in the high-mass region.
This explains why the branching fraction
of $H^\pm \to t b$ is reduced in
these regions. In the two lower-plots
shown in Fig.~\ref{Br1}, the decay
channels $H^\pm \to W^\pm h/H$ are
displayed, and they become the dominant
contributions as increasing 
of the charged Higgs masses.
%%%%%%%%%%%%%%%%%%%%%%%%%%%%%%%%%%%%%%%
%%%%%%%%%%%%%%%%%%%%%%%%%%%%%%%%%%%%%%%

\begin{figure}[]
\centering
\begin{tabular}{cc}
\includegraphics[width=8cm, height=5cm]
{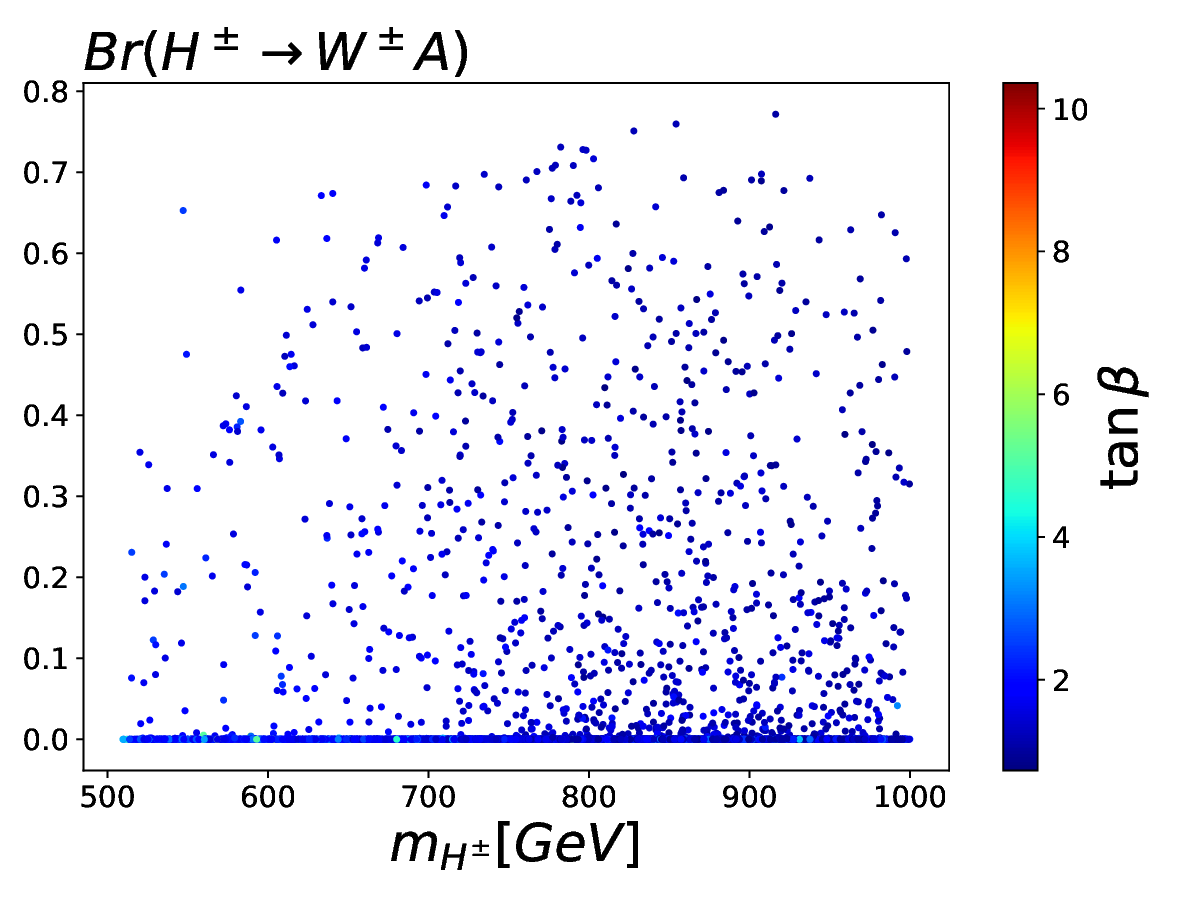}
&
\includegraphics[width=8cm, height=5cm]
{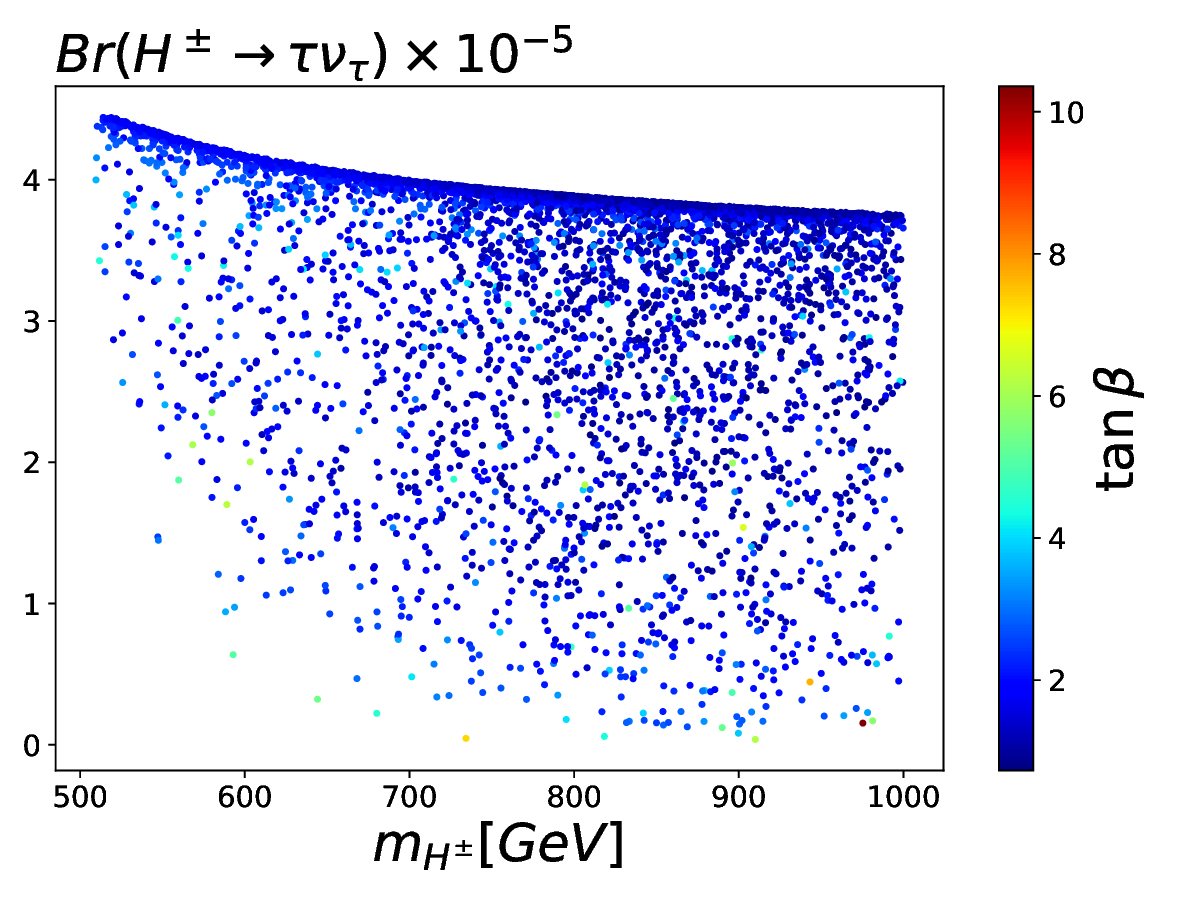}
\\
\includegraphics[width=8cm, height=5cm]
{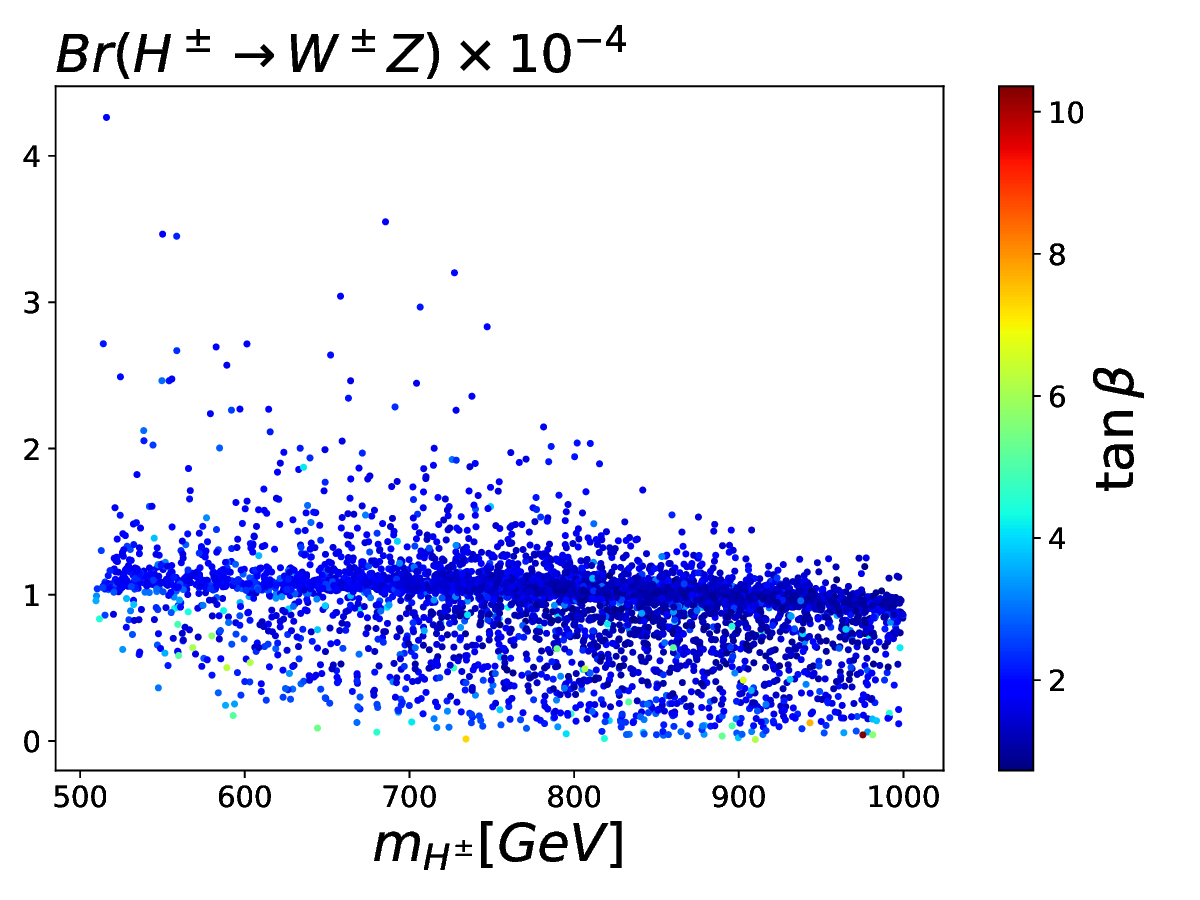}
&
\includegraphics[width=8cm, height=5cm]
{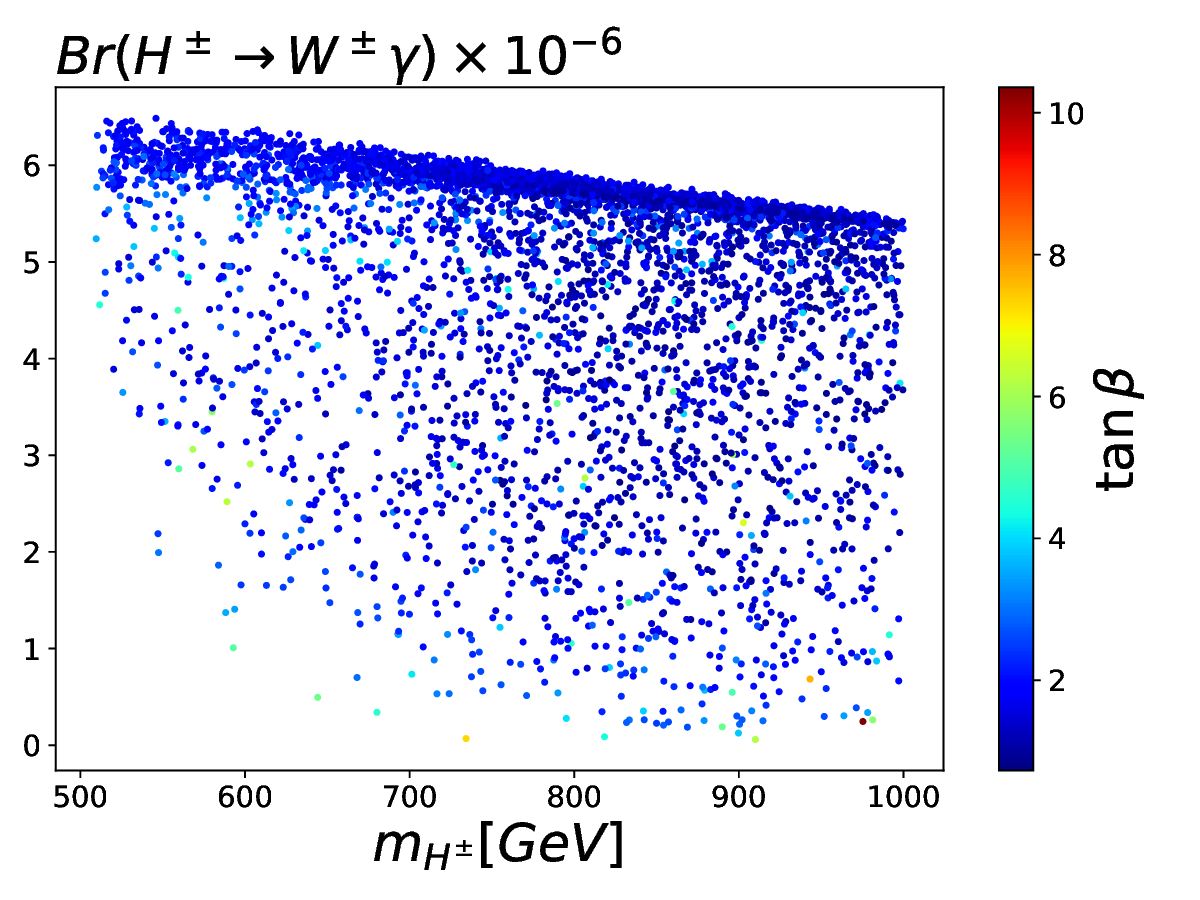}
%%%%%%%%%%%%%%%%%%%%%%%%%%%%%%%%%%%%%%%
\end{tabular}
\caption{\label{Br2}
The upper-left plot shows for the branching
fractions of the charged Higgs boson for the decay
$H^\pm \to W^\pm A$, and the upper-right plot shows
the decay rate of $H^\pm \to \tau \nu_{\tau}$.
Meanwhile, the lower-left plot corresponds to
$H^\pm \to W^\pm Z$ and the lower-right plot 
illustrates $H^\pm \to W^\pm \gamma$.
It is noted that all plots are generated as 
functions of the charged Higgs mass and $t_{\beta}$.
}
\end{figure}

In the above-left plot of Fig.~\ref{Br2},
the branching fraction of $H^\pm \to W^\pm A$
is shown as a function of $m_{H^\pm}$ and $\tan\beta$.
We observe behavior similar to that of the
$H^\pm \to W^\pm H$ channel.
All remaining decay channels of the charged Higgs, such as
$H^\pm \to \tau \nu_{\tau}$, $W^\pm Z$, and $W^\pm \gamma$,
are also shown in Fig.~\ref{Br2}, and we find that
their branching fractions are below $10^{-3}$
for all value of $t_{\beta}$ and charged Higgs masses.
%%%%%%%%%%%%%%%%%%%%%%%%%%%%%%%%%%%%%%%
%%%%%%%%%%%%%%%%%%%%%%%%%%%%%%%%%%%%%%%
In general, the results clearly indicate that the decay
modes $H^\pm \to t b$ and $H^\pm \to W^\pm \phi_j$
(with $\phi_j = h, H, A$) provide significant 
contributions.
In the scope of the current paper,
we focus on the $H^\pm \to t b$ channel
in the calculations of the signal significances
in the last subsection.
%%%%%%%%%%%%%%%%%%%%%%%%%%%%%%%%%%%%%%%
\subsection{Process 
$\mu^-\mu^+ \to H^\pm H^\mp h$ }
%%%%%%%%%%%%%%%%%%%%%%%%%%%%%%%%%%%%%%%
The process $\mu^- \mu^+ \to H^\pm H^\mp h$ is first examined
in this subsection. In the phenomenological analysis of this
production process, we take into consideration the dominant
decay channel of the charged Higgs boson, namely $H^\pm \to t b$.
Since the final-state top quarks subsequently decay into lighter
particles, we do not apply any cuts on the top quarks
themselves. Meanwhile, we apply the following cuts for
the final-state bottom and anti-bottom quarks as follows:
\begin{eqnarray}
p_{T}(b) \geq 20~\textrm{GeV}, \quad
p_{T}(\bar{b}) \geq 20~\textrm{GeV},
\quad 
|\eta(b)| \leq 2.4,
\quad 
|\eta(\bar{b})| \leq 2.4.
\end{eqnarray}
%%%%%%%%%%%%%%%%%%%%%%%%%%%%%%%%%%%%%%%
The cross sections (in ab) are shown
in the plane of the charged Higgs masses
and the mixing angle $\tan\beta$, as presented
in Fig.~\ref{signalSSh1}. The production
cross sections are displayed at
$\sqrt{s} = 3$ TeV in the left panel and
at $\sqrt{s} = 5$ TeV in the right panel.
For small values of $\tan\beta$, the cross
section is of the order of $\mathcal{O}(100)$ ab.
With the high integrated luminosities expected
at future multi-TeV colliders
(up to $\mathcal{L} = 3000$ fb$^{-1}$),
one expects to collect roughly $\sim \mathcal{O}(100)$ 
events. The same data are scanned over the charged Higgs
masses and the CP-even Higgs mass.
The cross sections are of the same order
as in the previous plots. The enhanced cross
sections occur primarily in the region
where $m_H \sim m_{H^\pm}$. 
%%%%%%%%%%%%%%%%%%%%%%%%%%%%%%%%%%%%%%%
\begin{figure}[]
\centering
\begin{tabular}{cc}
\includegraphics[width=8.3cm, height=7cm]
{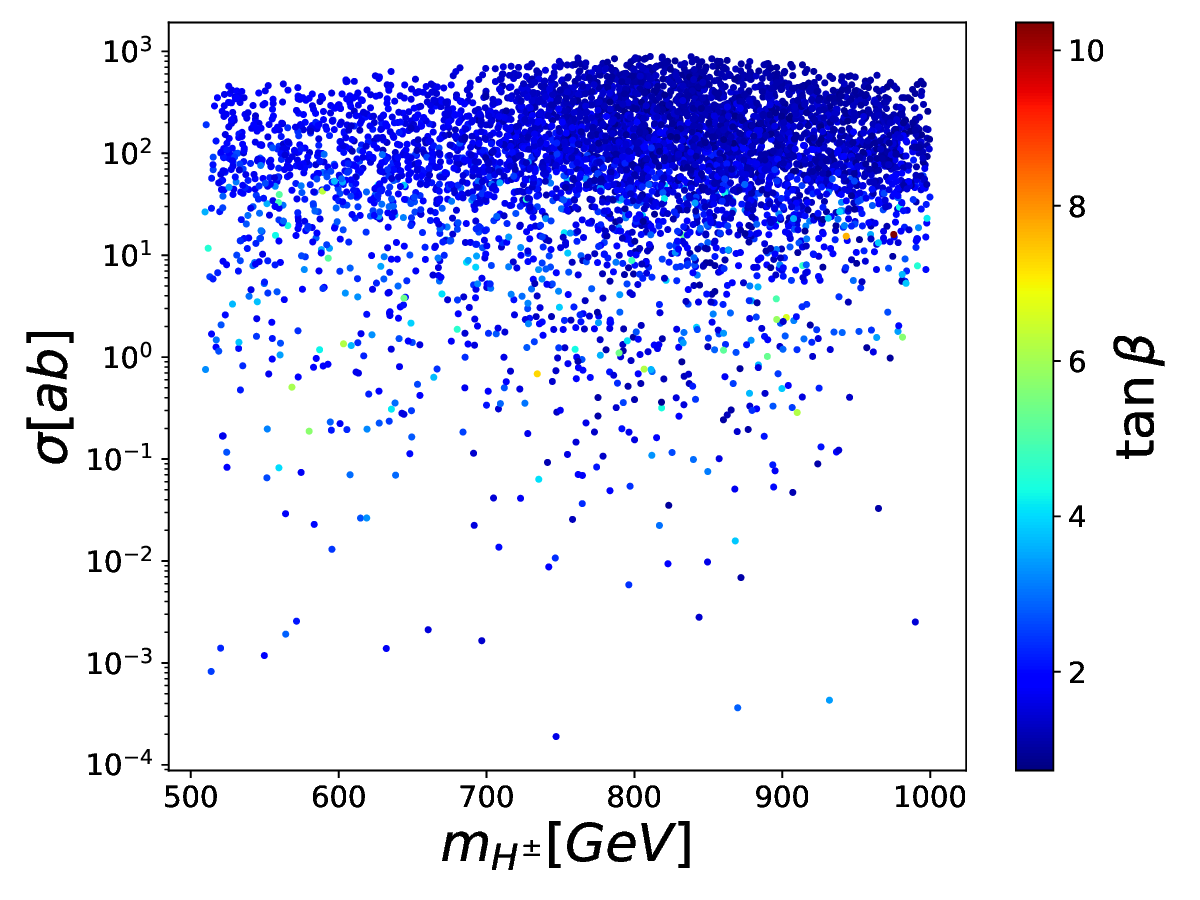}
&
\includegraphics[width=8.3cm, height=7cm]
{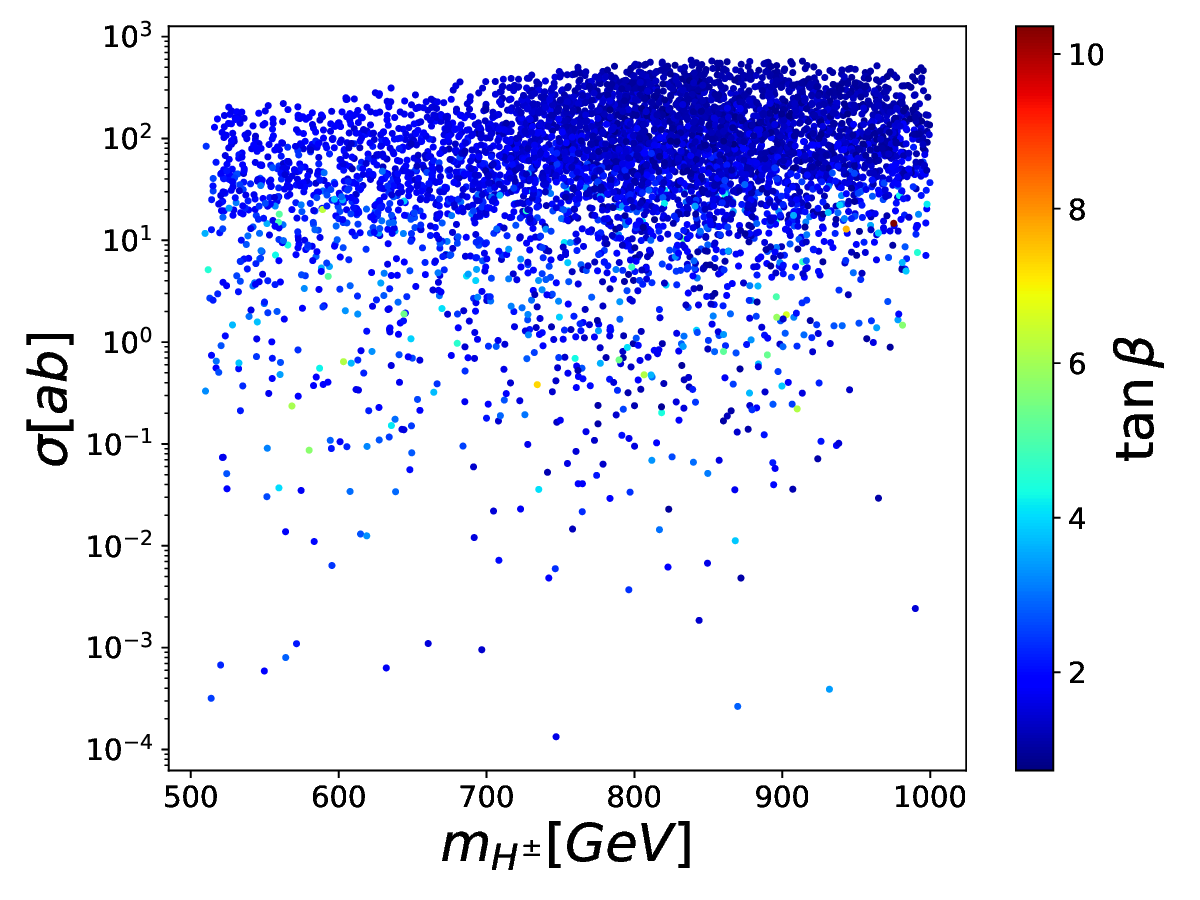}
\\
\includegraphics[width=8.3cm, height=7cm]
{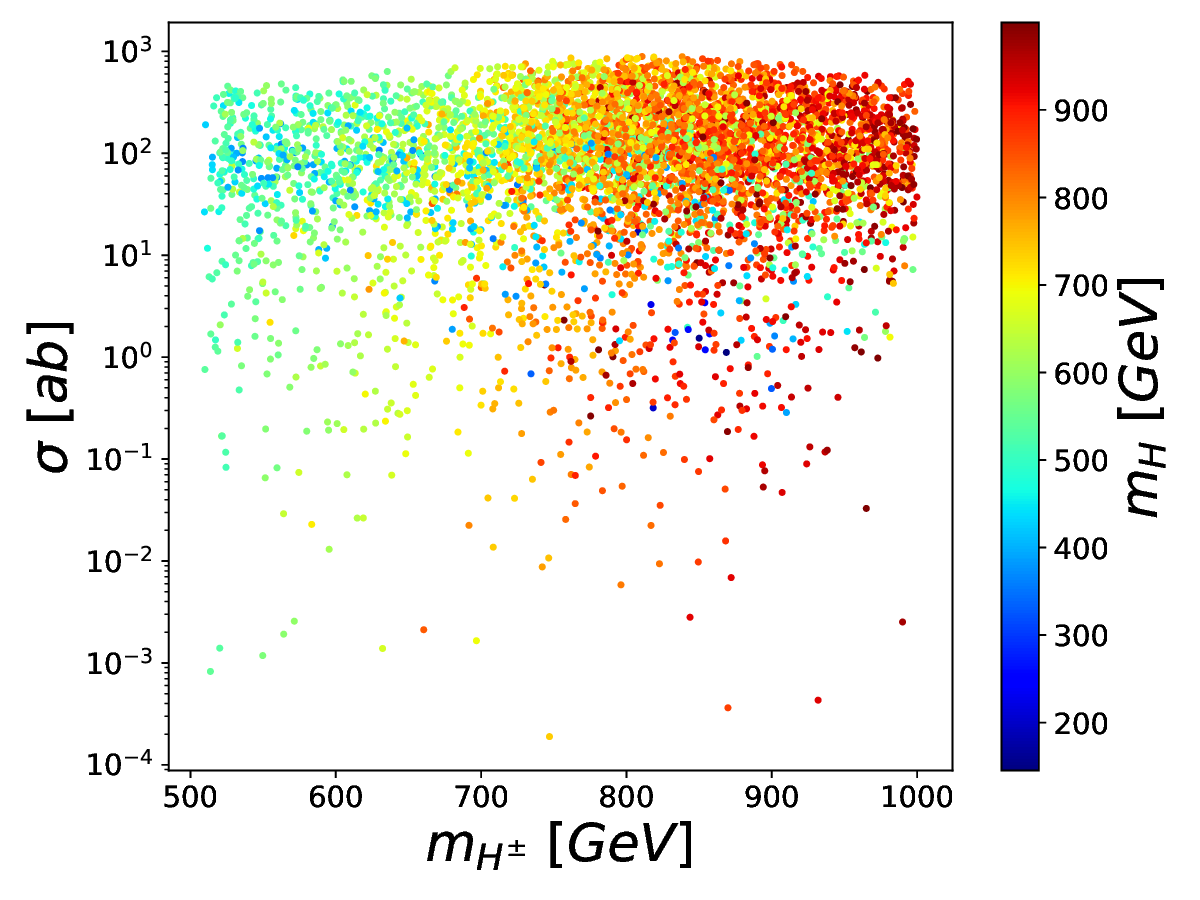}
&
\includegraphics[width=8.3cm, height=7cm]
{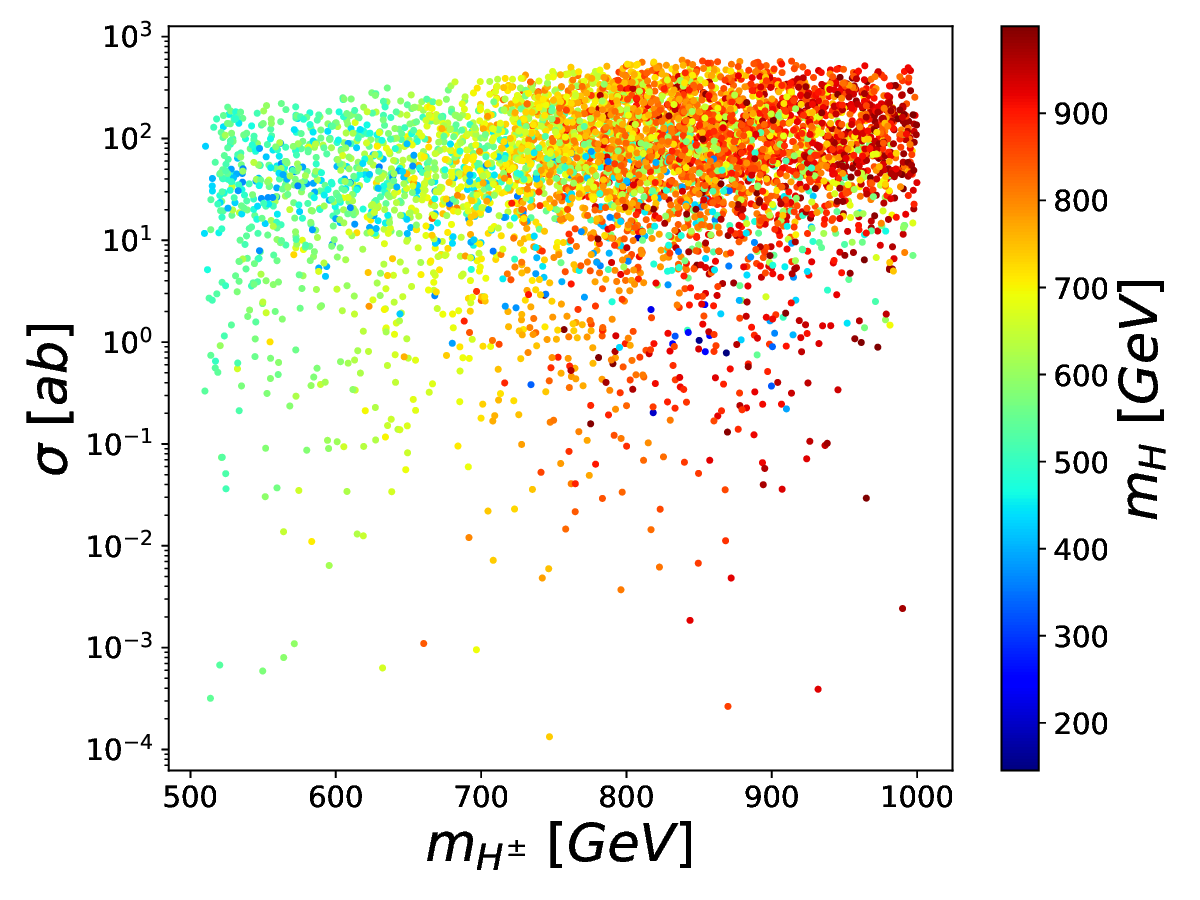}
\\
\includegraphics[width=8.3cm, height=7cm]
{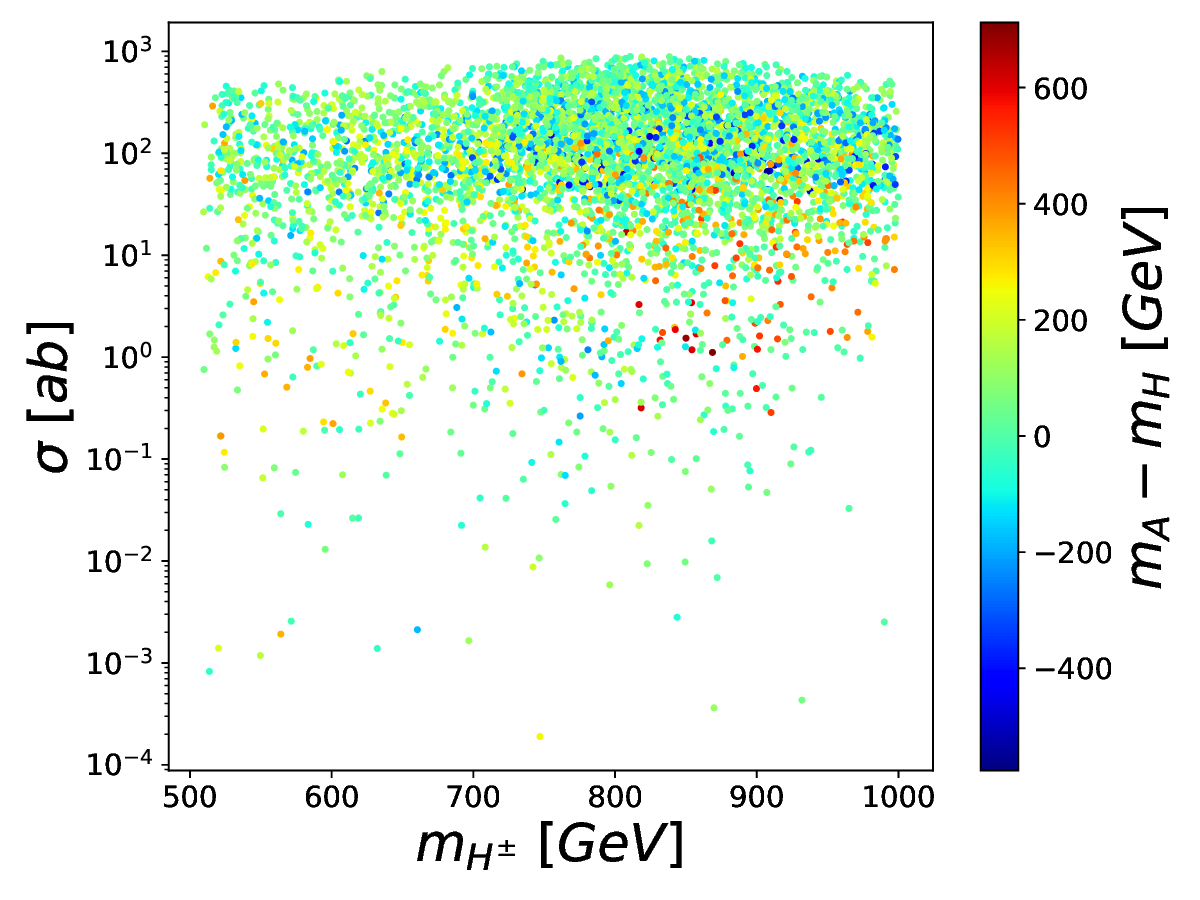}
&
\includegraphics[width=8.3cm, height=7cm]
{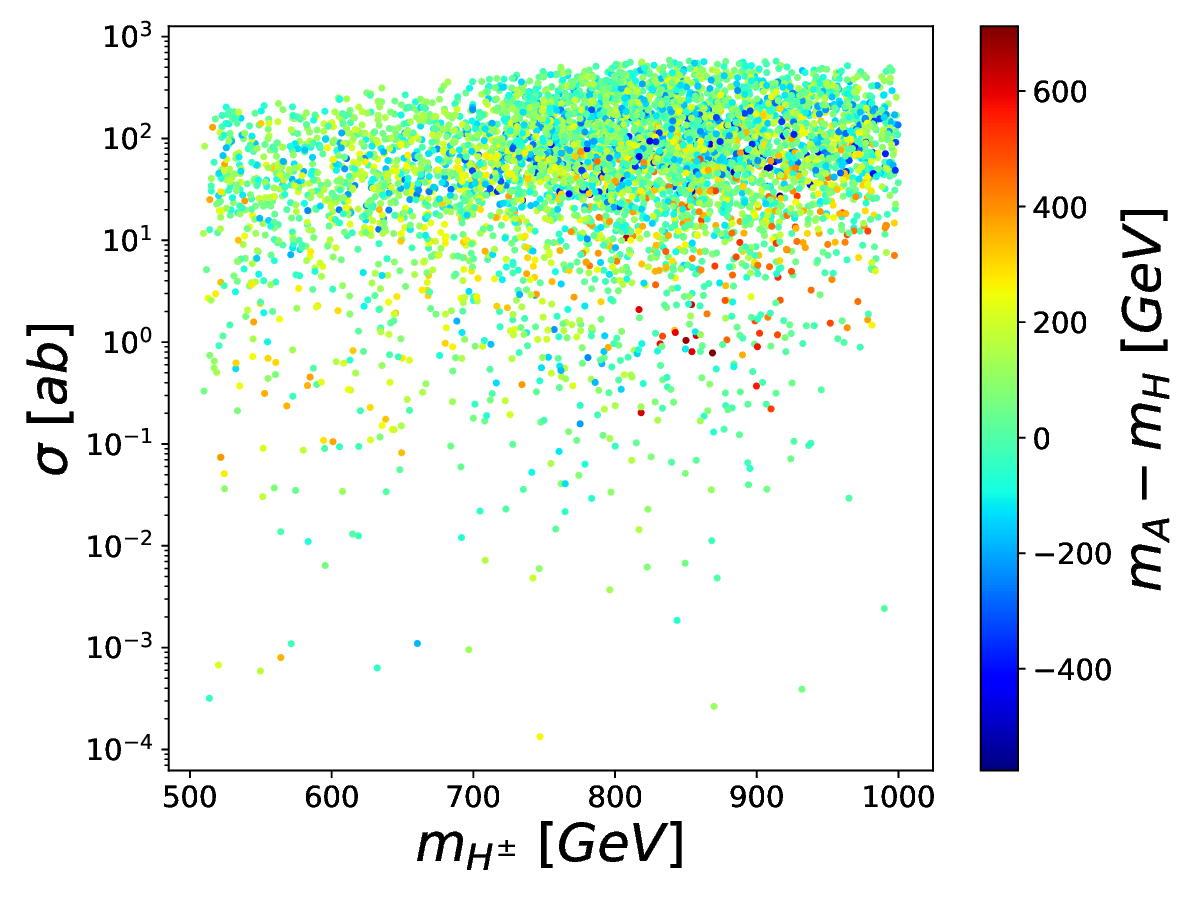}
%%%%%%%%%%%%%%%%%%%%%%%%%%%%%%%%%%%%%%%
\end{tabular}
\caption{\label{signalSSh1}
The production cross sections 
for $\mu^-\mu^+ \to H^\pm H^\mp h \to 
t\bar{t} b \bar{b} h$
at $\sqrt{s} = 3$~TeV (left) and 
$\sqrt{s} = 5$~TeV (right) 
are shown in the parameter space of 
$m_{H^\pm}$ and $t_{\beta}$ (upper plots), 
$m_{H^\pm}$ and $m_H$ (middle plots), 
and $m_{H^\pm}$ and $m_A - m_H$ 
(lower plots).}
\end{figure}
%%%%%%%%%%%%%%%%%%%%%%%%%%%%%%%%%%%%%%%
The data shows that the production cross
section reaches large values in the region of
$\tan\beta \lesssim 3$, and the mass difference
between the two scalar particles should be small
($|\Delta m_{\phi}| \lesssim 50$ GeV). 
There are two distinct regions for the 
CP-even Higgs where the production 
cross sections are enhanced. The first region
corresponds to $m_H \sim 500~\text{GeV}$ 
when $m_{H^\pm} \leq 750~\text{GeV}$. 
The second region appears for 
$m_H \geq 800~\text{GeV}$ when 
$m_{H^\pm} \geq 750~\text{GeV}$.
These scans provides
useful information for selecting benchmark points
to simulate the signal significances in
the last subsection.
%%%%%%%%%%%%%%%%%%%%%%%%%%%%%%%%%%%%%%%
\subsection{Process 
$\mu^-\mu^+ \to H^\pm H^\mp H$ }
\begin{figure}[]
\centering
\begin{tabular}{cc}
\includegraphics[width=8.3cm, height=7cm]
{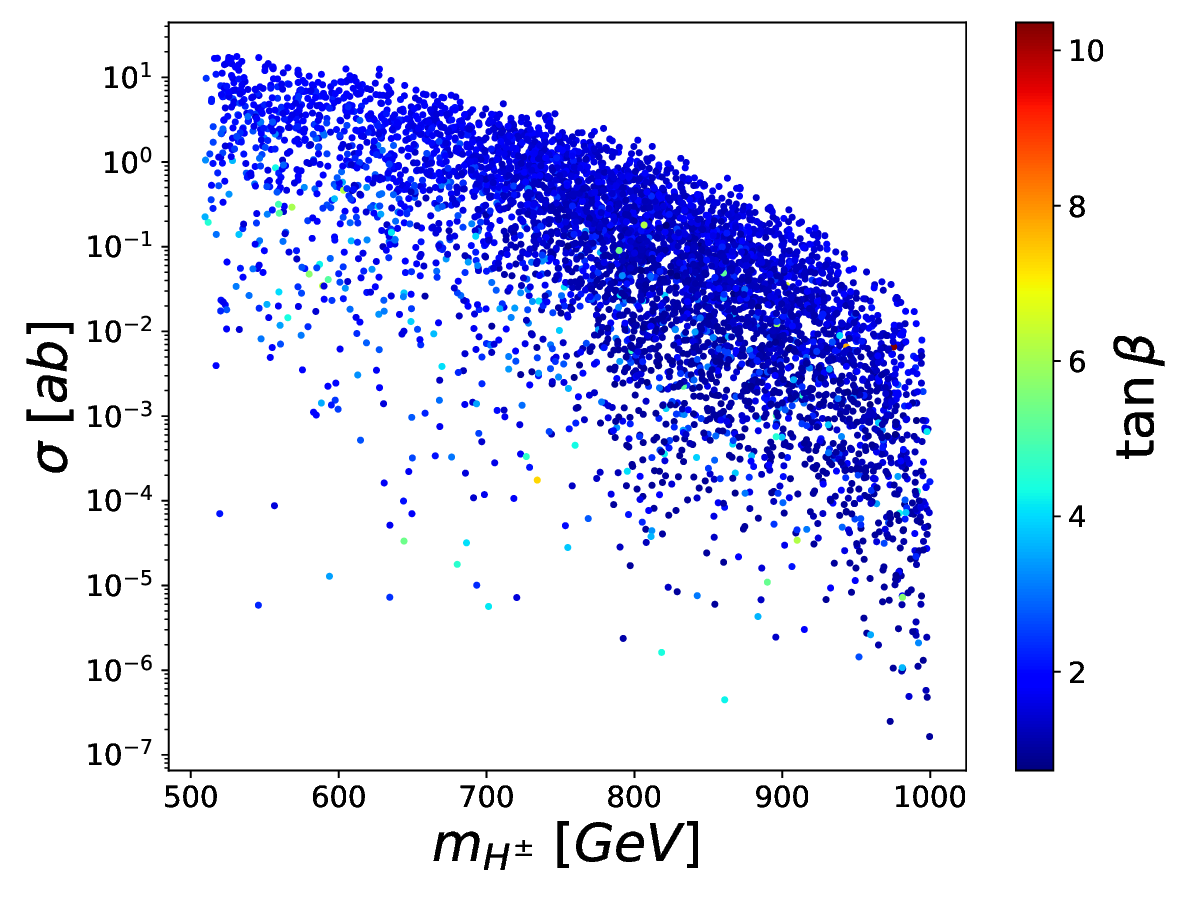}
&
\includegraphics[width=8.3cm, height=7cm]
{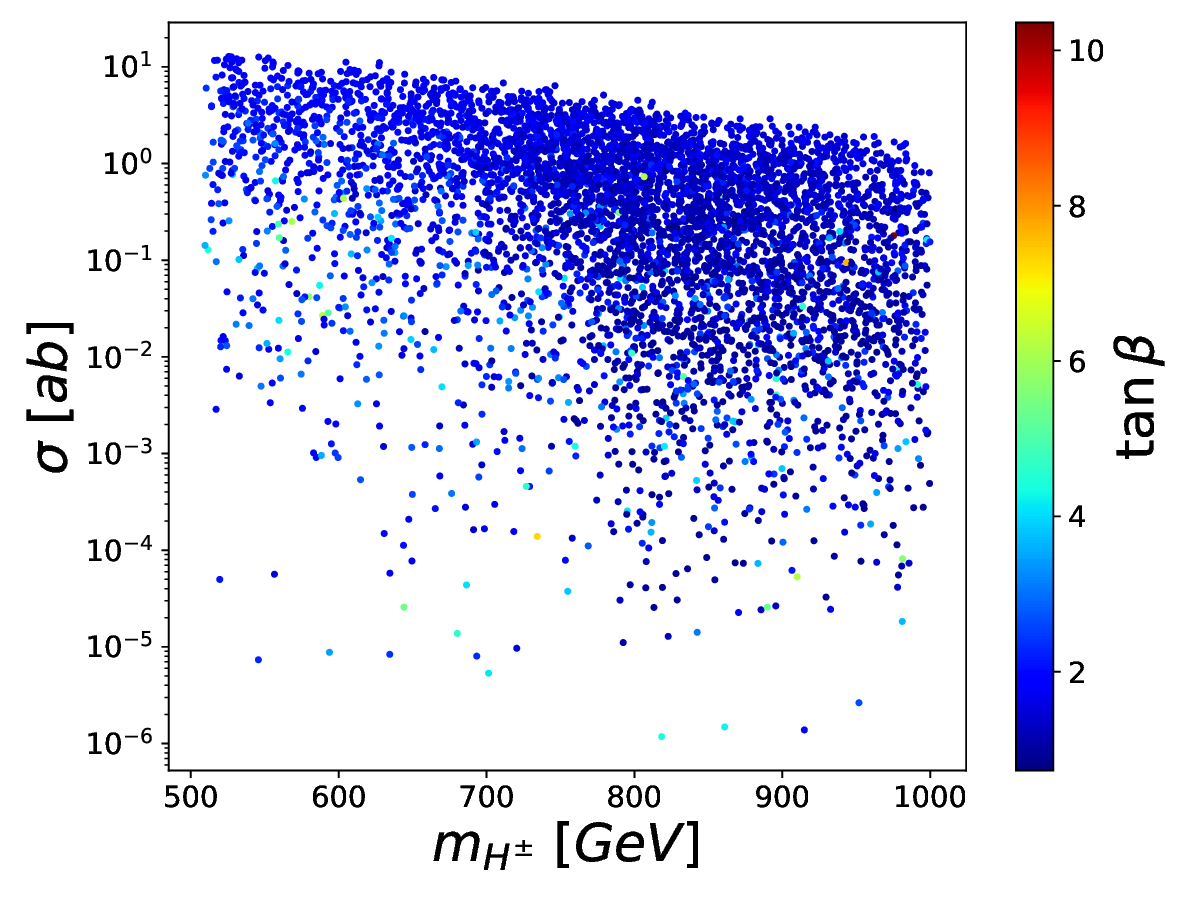}
\\
\includegraphics[width=8.3cm, height=7cm]
{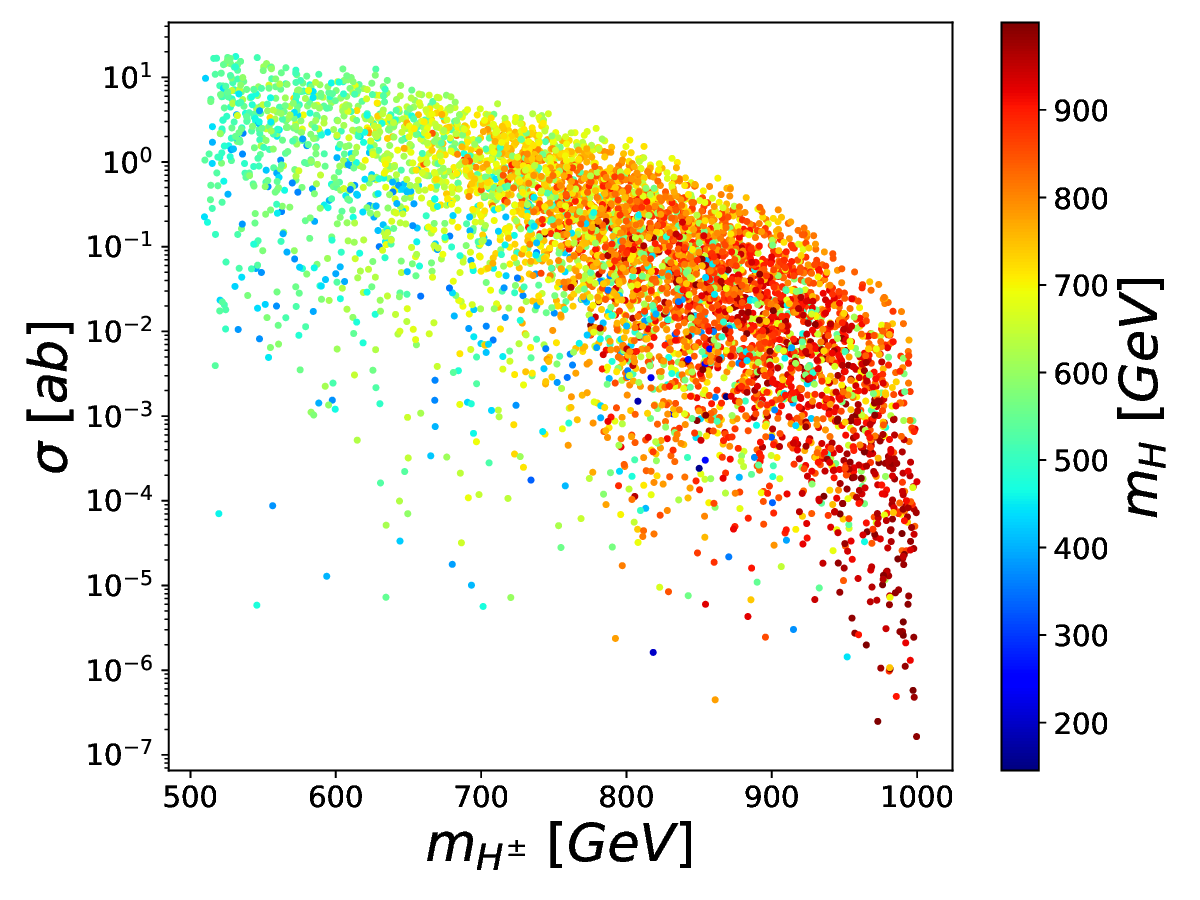}
&
\includegraphics[width=8.3cm, height=7cm]
{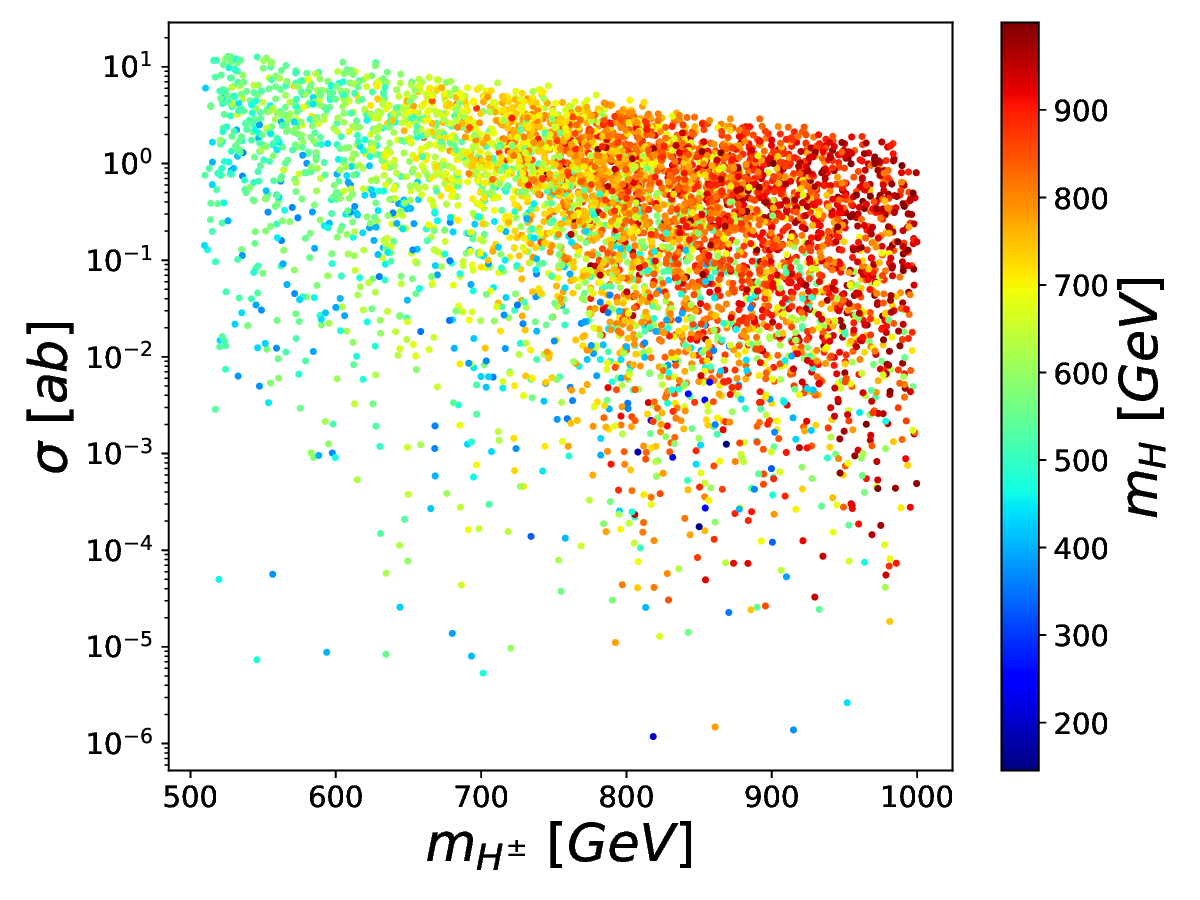}
\\
\includegraphics[width=8cm, height=7cm]
{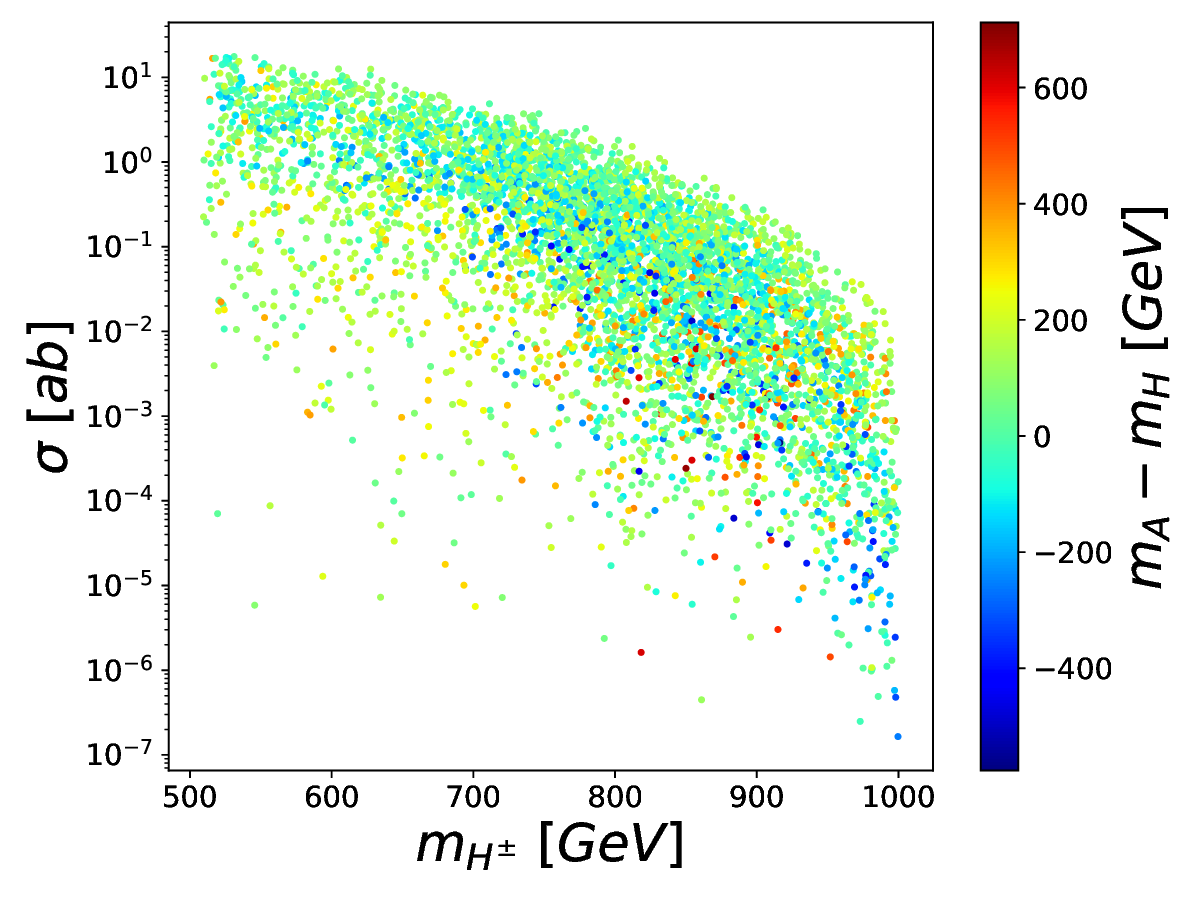}
&
\includegraphics[width=8cm, height=7cm]
{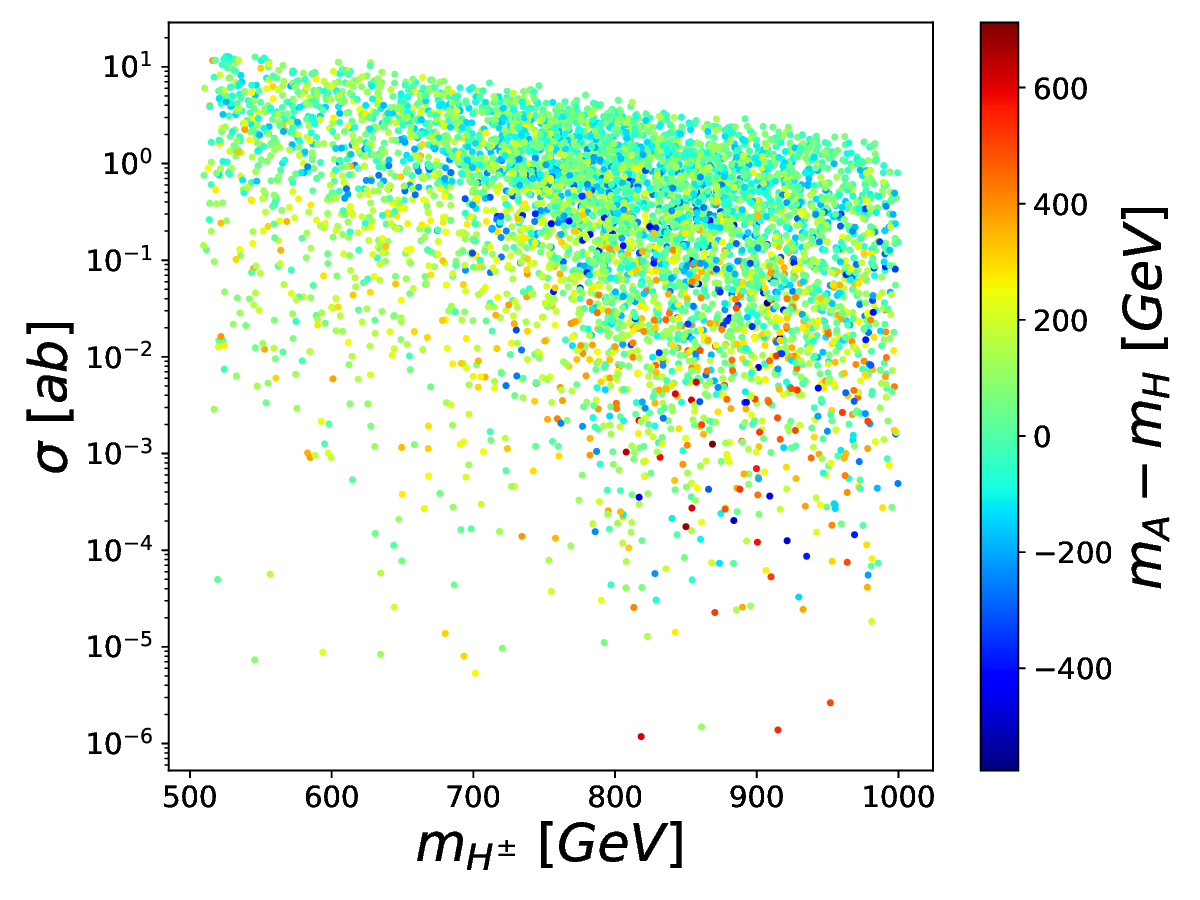}
%%%%%%%%%%%%%%%%%%%%%%%%%%%%%%%%%%%%%%%
\end{tabular}
\caption{\label{signalSSH1}
The production cross sections 
for $\mu^-\mu^+ \to H^\pm H^\mp H \to t\bar{t} b \bar{b} H$ at 
$\sqrt{s} = 3$~TeV (left) and $\sqrt{s} = 5$~TeV (right) 
are shown in the parameter space of 
$m_{H^\pm}$ and $t_{\beta}$ (upper plots), 
$m_{H^\pm}$ and $m_H$ (middle plots), 
and $m_{H^\pm}$ and $m_A - m_H$ (lower plots).
}
\end{figure}

%%%%%%%%%%%%%%%%%%%%%%%%%%%%%%%%%%%%%%%
We next examine the production process
$\mu^- \mu^+ \to H^\pm H^\mp H \to t\bar{t} b\bar{b} H$
at future multi-TeV muon colliders. We first scan
the production cross sections over the parameter
space of $m_{H^\pm}$ and $\tan\beta$, as
shown in the two upper plots of Fig.~\ref{signalSSH1}.
In particular, the left plot shows the cross sections
at a $3$ TeV of center-of-mass energy, and the right plot
is for a $5$ TeV of center-of-mass energy. Within the
entire range of $\sim 1 \leq t_{\beta} \leq 10$, 
the cross sections decrease with increasing charged 
Higgs masses. Due to the fact that the branching 
ratio of $H^\pm \to tb$ is dominant in the region 
of small values of $t_{\beta}$, we find that the 
cross sections focus mainly on the small 
values of $t_{\beta}$.
Meanwhile, the production cross sections are
presented in terms of $m_{H^\pm}$ and $m_H$
in the two middle plots of Fig.~\ref{signalSSH1}.
The left plot shows the cross section at 
$3$ TeV of center-of-mass energy, and the right plot
is for a $5$ TeV of center-of-mass energy.
The cross sections decrease as the
charged Higgs and CP-even Higgs masses increase.
The cross sections focus mainly on two regions of 
the CP-even Higgs. The first regime is
$m_{H^\pm} \leq 750$ GeV and $m_H \sim 500$ GeV, 
and the second one is $m_{H^\pm} \geq 750$ GeV and 
$m_H \geq 800$ GeV.
Finally, the same quantities are plotted 
as a function of $m_{H^\pm}$ and $m_A - m_H$
(at $3$ TeV for the left panel and at $5$ TeV 
for the right panel). 
The cross sections mainly focus on the region 
$|\Delta m_{\phi}|=|m_A-m_H| \lesssim 50$ GeV.
From the data,
we find that the production cross section reaches 
large values in the region of 
$\tan\beta \lesssim 3$, where the mass difference
between the two scalar particles is small
($|\Delta m_{\phi}| \lesssim 50$ GeV) and the 
charged Higgs mass is low ($m_{H^\pm} \leq 600$ GeV). 
In other regions the cross sections become negligible.
%%%%%%%%%%%%%%%%%%%%%%%%%%%%%%%%%%%%%%%
%%%%%%%%%%%%%%%%%%%%%%%%%%%%%%%%
\subsection{Significances} %%%%%
\begin{figure}[]
\centering
\begin{tabular}{cc}
\includegraphics[width=8.3cm, height=7cm]
{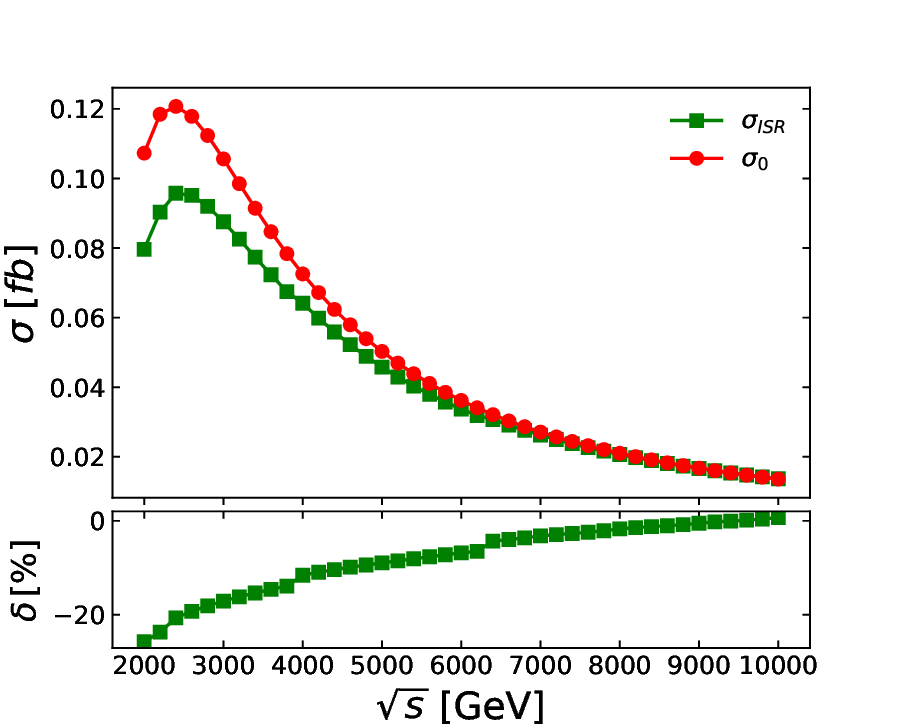}
&
\includegraphics[width=8.3cm, height=7cm]
{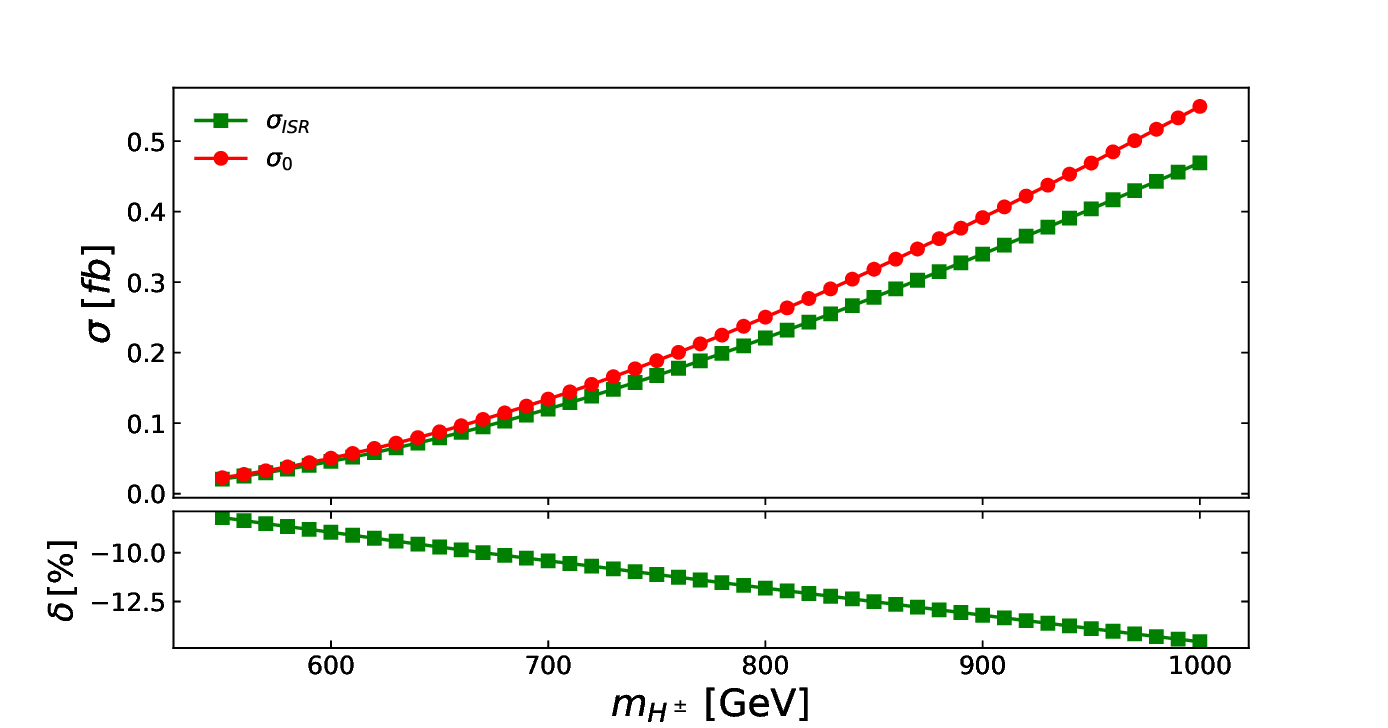}
%%%%%%%%%%%%%%%%%%%%%%%%%%%%%%%%%%%%%%%
\end{tabular}
\caption{\label{isrSqrt}
The ISR corrections to the process
$\mu^-\mu^+ \to H^\pm H^\mp h$ are 
studied at different center-of-mass energies 
$\sqrt{s}=2$ TeV to $10$ TeV (left)
and at fixing $\sqrt{s} = 5~\text{TeV}$ 
with varying charged Higgs masses 
$m_{H^\pm}=600$ GeV to $1000$ GeV (right).
}
\end{figure}

%%%%%%%%%%%%%%%%%%%%%%%%%%%%%%%%
In this subsection, the significances for 
$\mu^- \mu^+ \to H^\pm H^\mp h$ are computed 
at several selected benchmark points. Before presenting 
the results, we note that initial-state radiation (ISR) 
is important at future lepton colliders, 
as reported in many of our previous works. 
Therefore, we first consider the two-loop ISR 
corrections to the process 
$\mu^- \mu^+ \to H^\pm H^\mp h$ as performed
in Ref.~\cite{Phan:2025pjt}. 
Detailed formulas for evaluating the ISR corrections 
are provided in Ref.~\cite{Phan:2025pjt}. 
In this subsection, we first present the effects of ISR 
corrections on the production process. In the left plot of
Fig.~\ref{isrSqrt}, the ISR corrections are computed at 
center-of-mass energies ranging from $2$~TeV to $10$~TeV, 
taking the process $\mu^- \mu^+ \to H^\pm H^\mp h$ as a 
representative example for our survey.
For this study, we use BP1 as the input parameters. 
In the right plot of Fig.~\ref{isrSqrt}, we show 
the ISR corrections at 
$\sqrt{s} = 5~\text{TeV}$ with varying charged Higgs 
masses. We find that the ISR corrections range from 
$-25\%$ to $-1\%$ for all center-of-mass 
energies from $2$~TeV to $10$~TeV, while 
the ISR corrections range from 
$-4\%$ to $-15\%$ for all charged Higgs masses
at $\sqrt{s}=5$ TeV.
%%%%%%%%%%%%%%%%%%%%%%%%%%%%%%%%%%%%%%%
%%%%%%%%%%%%%%%%%%%%%%%%%%%%%%%%%%%%%%%
%%%%%%%%%%%%%%%%%%%%%%%%%%%%%%%%%%%%%%%
\begin{table}[]
\centering
\begin{tabular}{lcccccc}
\hline\hline
Benchmark & $m_H$ (GeV) & $m_A$ (GeV) & $t_\beta$ 
& $s_{\beta-\alpha}$ 
& $m_{12}^2$ (GeV$^2$) & $m_{H^\pm}$ (GeV) \\
\hline
BP1 & $500$ & $500$ & $1.5$ & $0.99$ & $10^{5}$ & $550$--$1000$ \\
BP2 & $800$ & $800$ & $1.5$ & $0.99$ & $10^{5}$ & $550$--$1000$ \\
\hline\hline
\end{tabular}
\caption{\label{BP} Benchmark points selected 
in the analysis. }
\end{table}
%%%%%%%%%%%%%%%%%%%%%%%%%%%%%%%%%%%%%%%

In this section, we compute the signal 
significances for the process 
$\mu^- \mu^+ \to H^\pm H^\mp h  
\to t\bar{t}\, b\bar{b}\, h 
\to t\bar{t}\, b\bar{b}\, b\bar{b}$ 
evaluated at several selected benchmark points above 
for future multi--TeV muon colliders. The benchmark 
points in the allowed parameter space 
are shown in the Table~\ref{BP}.
For the SM background, we consider 
two processes like
$\mu^- \mu^+ \to t\bar{t}\, b\bar{b}\, h 
\to t\bar{t}\, b\bar{b}\, b\bar{b}$ and 
$\mu^- \mu^+ \to t\bar{t}\, b\bar{b}\, Z 
\to t\bar{t}\, b\bar{b}\, b\bar{b} $. 
The background processes 
are calculated with the help of 
{\tt MadGraph5MC@NLO}~\cite{Frederix:2018nkq}.  
The significances are then computed as follows:
%%%%%%%%%%%%%%%%%%%%%%%%%%%%%%%%%%%%%
\begin{eqnarray}
 \mathcal{S} = \frac{N_{\textrm{S} }}
 {
 \sqrt{
 N_{\textrm{S} }
 + \varepsilon_B N_{\textrm{B} }
 }
 }.
\end{eqnarray}
%%%%%%%%%%%%%%%%%%%%%%%%%%%%%%%%%%%%%
Where $N_{\textrm{S/B}} = \mathcal{L}
\sigma_{\textrm{S/B}}$ 
is corresponding for number of event and 
background. The factor $\varepsilon_B$ is 
for systematic uncertainty fraction on the
background yield. 
We take $\varepsilon_B=1, 1.5$ in this work.

For both the signal and the SM backgrounds, 
we apply the following cuts on the final-state 
bottom and anti-bottom quarks originating from 
the charged Higgs decay as follows:
%%%%%%%%%%%%%%%%%%%%%%%%%%%%%%%%%%%%%%%
\begin{eqnarray}
 p_{T}(b) \geq 20~\textrm{GeV},
 \quad p_{T}(\bar{b}) \geq 20~\textrm{GeV}, 
 \quad 
 |\eta(b)| \leq 2.4,\quad 
 |\eta(\bar{b})| \leq 2.4.
\end{eqnarray}
%%%%%%%%%%%%%%%%%%%%%%%%%%%%%%%%%%%%%%%%%%%%%%
For the signal events, the decay $h \to b\bar{b}$ 
with a branching fraction of $53\%$ is taken 
from particle data 
group~\cite{ParticleDataGroup:2024cfk}.

In Fig.~\ref{signalSSh1},
the significances are compted at 
$\sqrt{s} = 3~\text{TeV}$, with BP1 shown 
on the left panel and BP2 on the right.  
In all plots, the blue, red, and green 
curves correspond to the integrated 
luminosities $\mathcal{L} = 0.5$, 
$1$, and $3~\text{ab}^{-1}$, respectively.  
The dashed lines represent the case 
$\varepsilon= 1$, while the solid lines 
correspond to $\varepsilon= 1.5$.
Our results indicate that the significances
can rearch $8$ for BP1 and $\sim 19$ for the 
second benchmark point BP2 at $\mathcal{L}=3$ 
ab$^{-1}$. 
%%%%%%%%%%%%%%%%%%%%%%%%%%%%%%%%%%%%%%%
\begin{figure}[]
\centering
\begin{tabular}{cc}
\includegraphics[width=8.3cm, height=6cm]
{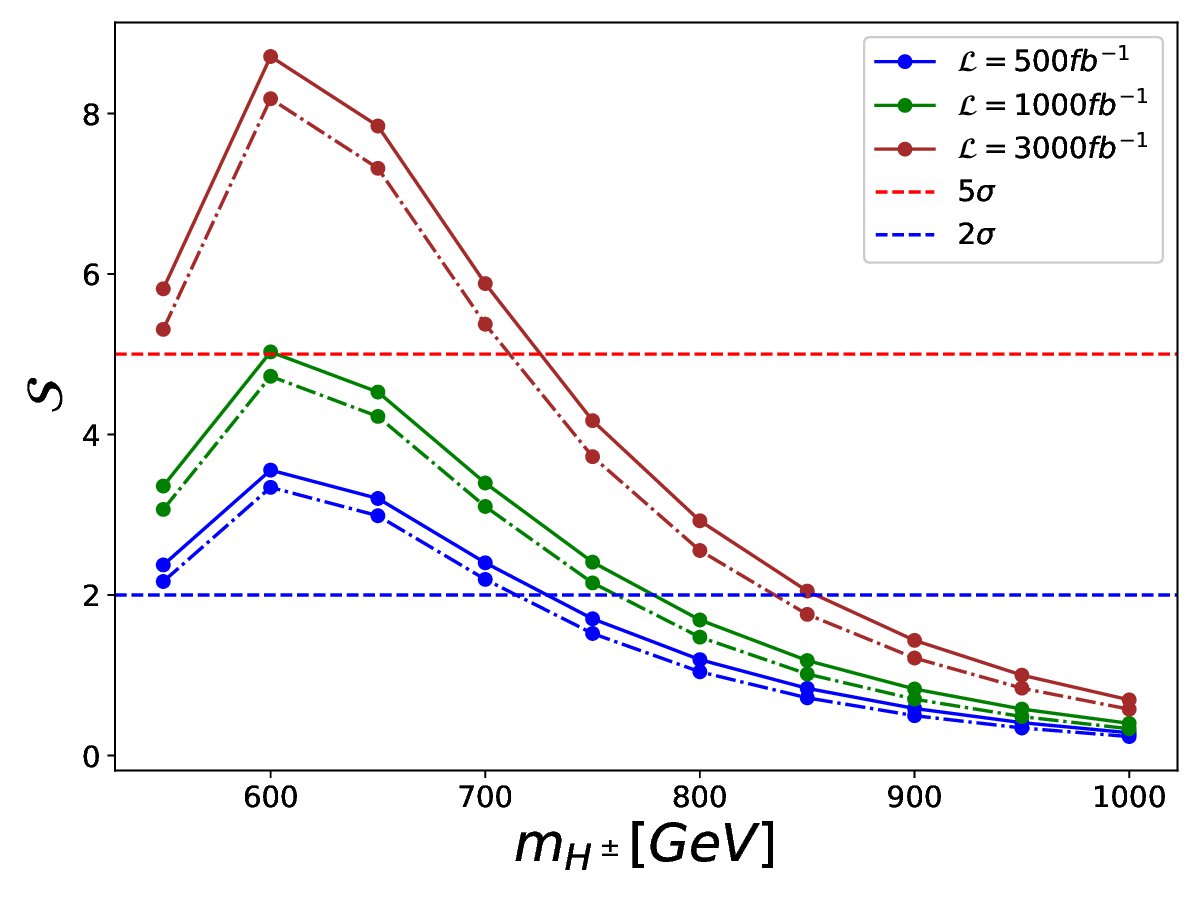}
&
\includegraphics[width=8.3cm, height=6cm]
{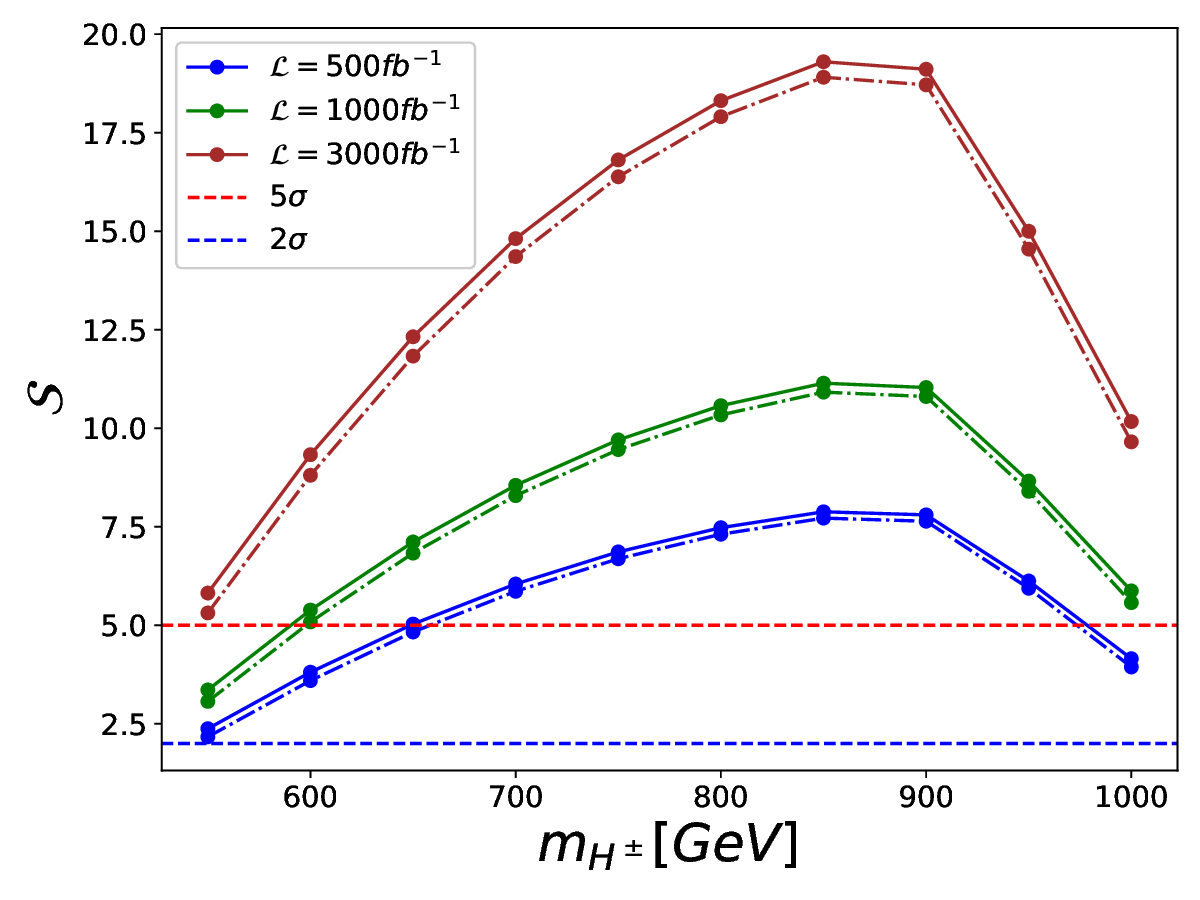}
%%%%%%%%%%%%%%%%%%%%%%%%%%%%%%%%%%%%%%%
\end{tabular}
\caption{\label{signalSSh1}
the significances are shown at $\sqrt{s} = 3~\text{TeV}$, 
with BP1 displayed on the left panel and BP2 on the right.}
\end{figure}
%%%%%%%%%%%%%%%%%%%%%%%%%%%%%%%%%%%%%%%

\begin{figure}[]
\centering
\begin{tabular}{cc}
\includegraphics[width=8.3cm, height=6cm]
{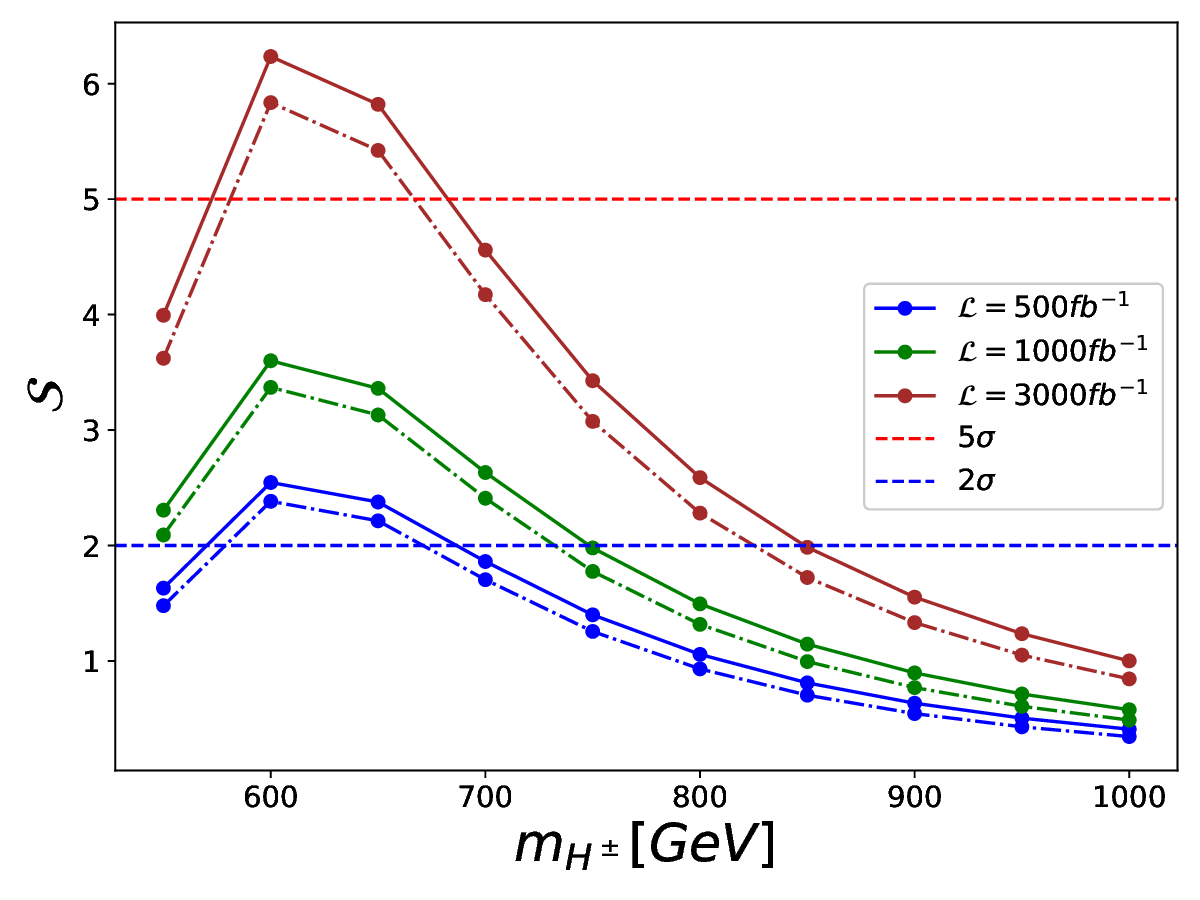}
&
\includegraphics[width=8.3cm, height=6cm]
{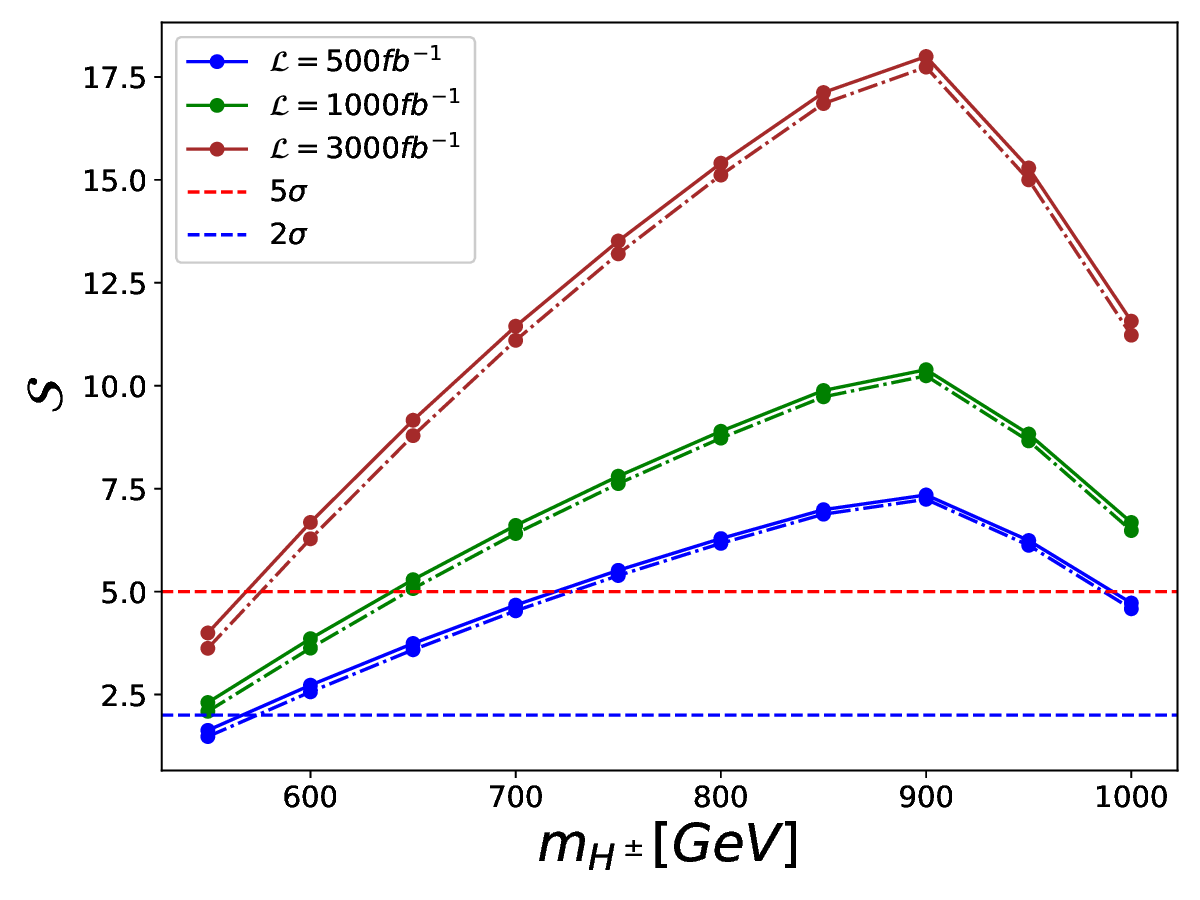}
%%%%%%%%%%%%%%%%%%%%%%%%%%%%%%%%%%%%%%%
\end{tabular}
\caption{\label{signalSSh2}
the significances are shown at $\sqrt{s} = 5~\text{TeV}$, 
with BP1 displayed on the left panel and BP2 on the right.}
\end{figure}

We next consider the significances
at $\sqrt{s} = 5~\text{TeV}$ in which 
the results for BP1 displaying
on the left panel and for BP2 on the right
of Fig.~\ref{signalSSh2}.  
In all plots, we use the same notation for the lines 
as previous Figures. It is observed that 
the significances rearch $6$ for BP1
and $\sim 17$ for the 
second benchmark point BP2 at $\mathcal{L}=3$ 
ab$^{-1}$. 
%%%%%%%%%%%%%%%%%%%%%%%%%%%%%%%%%%%%%%%
%%%%%%%%%%%%%%%%%%%%%%%%%%%%%%%%%%%%%%%

\begin{figure}[H]
\centering
\begin{tabular}{cc}
\includegraphics[width=8.3cm, height=6cm]
{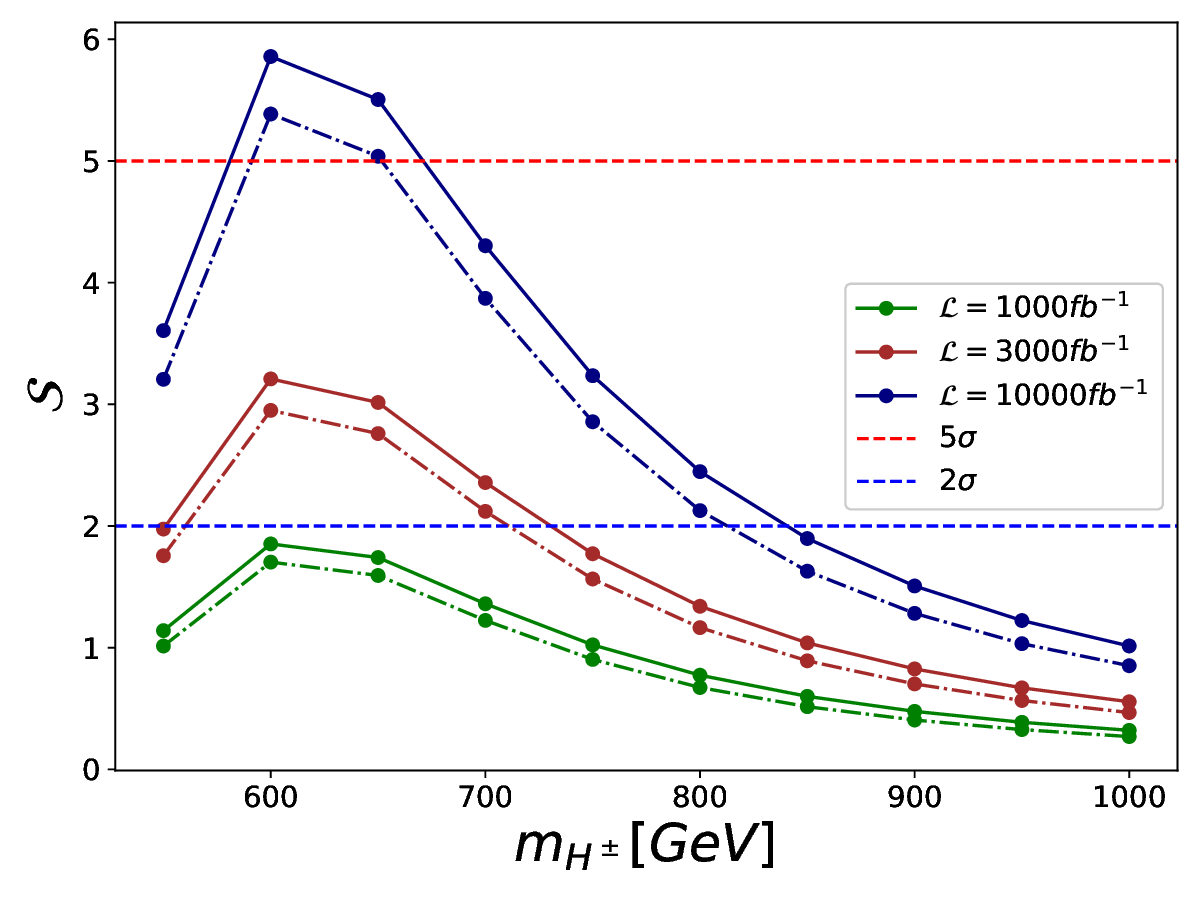}
&
\includegraphics[width=8.3cm, height=6cm]
{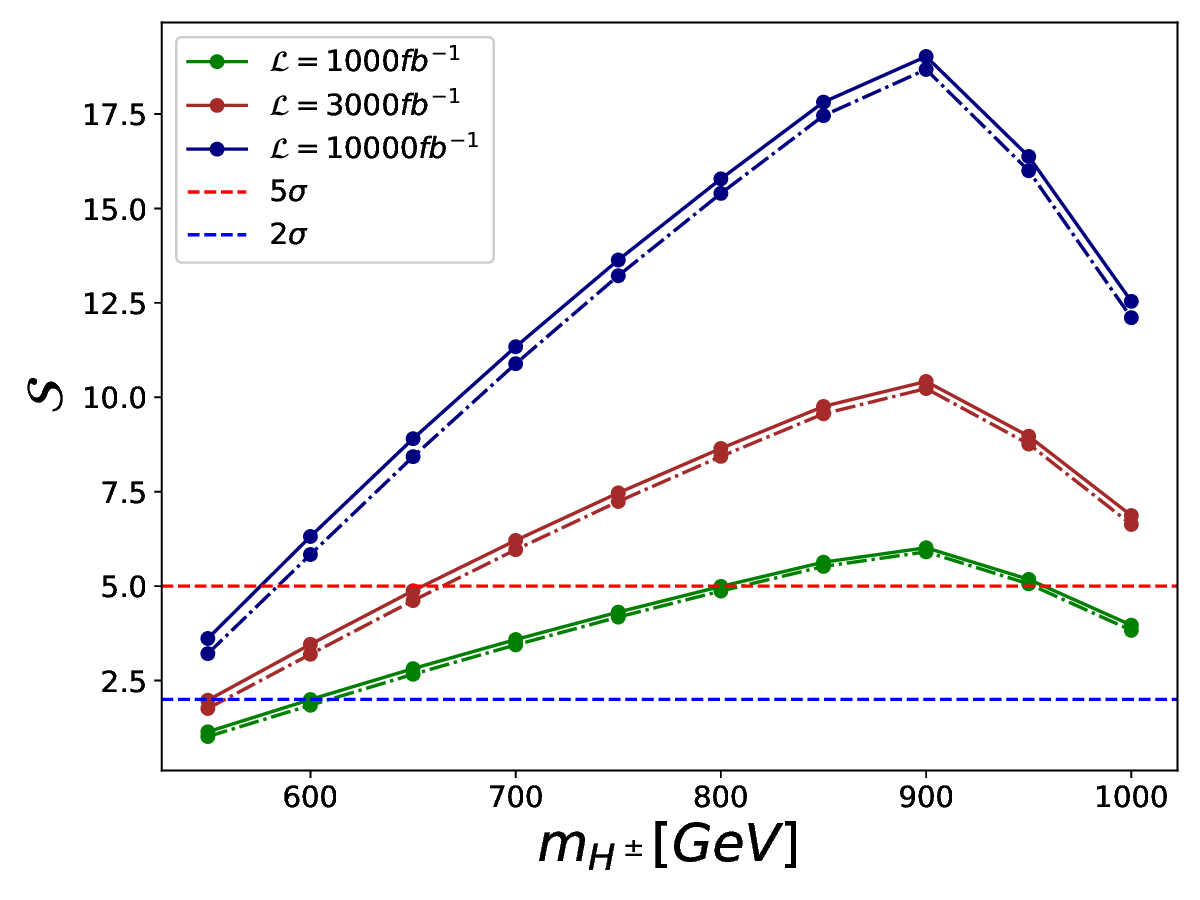}
%%%%%%%%%%%%%%%%%%%%%%%%%%%%%%%%%%%%%%%
\end{tabular}
\caption{\label{signalSSh3}
the significances are shown at $\sqrt{s} =10 ~\text{TeV}$, 
with BP1 displayed on the left panel and BP2 on the right.}
\end{figure}

Finally, the significances are also evaluated
at $\sqrt{s} = 10~\text{TeV}$, 
with BP1 displayed on the left panel 
and BP2 on the right of the scatter plots
~\ref{signalSSh3}. We obtain nearly
the same significances which are around $6$
for BP1 and around $17$ for BP2
at $\mathcal{L}=10$ab$^{-1}$. 
%%%%%%%%%%%%%%%%%%%%%%%%%%%%%%%%%%%%%%%
%%%%%%%%%%%%%%%%%%%%%%%%%%%%%%%%%%%%%%
We emphasize that the significances are enhanced
when $m_{H^\pm} \sim m_H + m_h$, as their signal cross sections
also peak under this condition, as shown in Fig.~\ref{signalSSH1}.
The peaks arise from the exchange of $s$-channel CP-even Higgses.
Furthermore, if we consider the top quark decay to 
leptons and bottom quark like 
$t \to \ell \bar{\nu}_{\ell} b$ with $\ell = e, \mu, \tau$,
using its branching ratio of $0.337$ as given in 
particle data group~\cite{ParticleDataGroup:2024cfk}, 
we then have six bottom quarks in the final state 
and apply a b-tagging factor of $\sim 0.8$. With these 
values, the significances for the process 
are scaled down by a factor of $\sqrt{0.337^2 
\times 0.8^6} \sim 0.2$.
Subsequently, the significance for BP1 will be reduced
around $\sim 1.3$, and for BP2 around $\sim 3.6$. 
In comparison with our previous
work~\cite{Phan:2025pjt}, these production processes 
in the Type-Y THDM are difficult to 
probe at multi-TeV muon colliders. Last but not least,
detector simulations for events with two leptons and 
six bottom quarks should be considered, as they also 
affect the significances. These topics will be 
addressed in our future work.
%%%%%%%%%%%%%%%%%%%%%%%%%%%
\section{Conclusions} %%%%%
%%%%%%%%%%%%%%%%%%%%%%%%%%%%%%%%%%%%%%%%%%%%%%%%%%%%%%%
We report the first results for charged Higgs boson
production associated with CP-even Higgs states at
future multi-TeV muon colliders in the scenario of
the Type-Y Two-Higgs-Doublet Model. The updated parameter
regions for the Type-Y Two-Higgs-Doublet
Model are obtained in this study. Based on the allowed
parameter space, we compute all two-body decay channels
of the charged Higgs boson with the help of {\tt H-COUP}.
The production processes
$\mu^- \mu^+ \to H^\pm H^\mp h/H
\to t\bar{t}b\bar{b}\; h/H$ are also scanned over 
the parameter space of the Type-Y THDM. In the allowed 
regions of the THDM, one can observe $\mathcal{O}(100)$ 
events, providing an opportunity to probe the charged 
Higgs bosons produced in association with the heavy 
CP-even Higgs at future multi-TeV muon colliders.
We compute the signal significances for
$\mu^- \mu^+ \to H^\pm H^\mp h \to t\bar{t} b\bar{b} h
\to t\bar{t} b\bar{b} b\bar{b}$ with respect to the
Standard Model background. We
also estimate the significances by including the top
quark decay into leptons and bottom quarks and 
considering $b$-tagging. Thanks to the high integrated
luminosities anticipated at future multi–TeV muon
colliders, we find that the signal significances 
can exceed $\sim 3\sigma$ at several benchmark points
selected in the viable parameter space of the Type-Y THDM.
%%%%%%%%%%%%%%%%%%%%%%%%%%%%%%%%%%%%%%%%%%
\\

\noindent
{\bf Acknowledgment:}~
This research is funded by Vietnam
National Foundation for Science and
Technology Development (NAFOSTED) under
the grant number $103.01$-$2023.16$.
%%%%%%%%%%%%%%%%%%%%%%%%%%%%%%%%%%%%%%%%%%%%
\subsection*{Process $\mu^+\mu^-
\to H^{\pm}H^{\mp}\phi_k$ for $\phi_k=h,~H$}
%%%%%%%%%%%%%%%%%%%%%%%%%%%%%%%%%%%%%%%%%%%%
In this appendix, we show a compact form for the
tree-level amplitude of the process $\mu^+ \mu^- 
\to H^{\pm} H^{\mp} \phi_k$ for $\phi_k = h,~H$ 
in the context of the THDM. The results are computed 
with considering all Feynman diagrams 
with 't Hooft-Feynman gauge. Amplitude for 
processes $\mu^+\mu^-
\to H^{\pm}H^{\mp} h$ 
are given: 
\begin{eqnarray}
i\mathcal{M} 
&= \sum\limits_{k = 1}^{8} F_k \langle v|P_k|u \rangle,
\end{eqnarray}
where the tree-level form factors $F_k$ are expressed
in terms of the following kinematic variables such as
$s =s_{12}= \left(k_1 + k_2\right)^2$,
$s_{34} = \left(k_3 + k_4\right)^2$,
$t= t_{13}= \left(k_1 - k_3\right)^2$,
$t_{14} = \left(k_1 - k_4\right)^2$,
$t_{24} = \left(k_2 - k_4\right)^2$,
and $u=u_{23}= \left(k_2 - k_3\right)^2$. 
The explicit expressions for the 
form factors $F_k$ are given by:
%%%%%%%%%%%%%%%%%%%%%%%%%%%%%%%%%%%
\begin{eqnarray}
F_1 &=& 
g_{h f f}^R 
\dfrac{g_{H^\pm H^\mp h h}
}{s - m_h^2}
+
g_{Hff}^R 
\dfrac{g_{H^\pm H^\mp H h} 
} {s - m_H^2}
+
\dfrac{g_{h H^\pm H^\mp}}
{s_{34} - m_{H^\pm}^2}
\left[
g_{h f f}^R
\dfrac{g_{h H^\pm H^\mp} }{s - m_h^2}
+
g_{H f f}^R
\dfrac{g_{H H^\pm H^\mp} }{s - m_H^2}
\right]
\nonumber\\
&& + 
\dfrac{m_\mu\;  g_{h H^\pm H^\mp} 
\; g_{H^\pm H^\mp V}
\cdot 
g_{V f f}^{R-L}  }
{\left(s -m_Z^2\right)
\left(s_{34} - m_{H^\pm}^2\right)}
+ 
\dfrac{\pi \alpha c_{\beta -\alpha}}
{s_W^2 \left(s_{34} - m_W^2\right)}
\left[
C_8
\cdot
g_{H f f}^R
\dfrac{s_{\beta - \alpha} 
}{s - m_H^2}
-
C_5
\cdot 
g_{h f f}^R
\dfrac{c_{\beta - \alpha} 
}{s - m_h^2}
\right]
\nonumber\\
%%%%%%%%%%%%%%%%%%%%%%%%%%%
&& 
+ 
\dfrac{g_{h H^\pm H^\mp} }
{3 m_{H^\pm}^2 + 2 m_\mu^2 
- s_{34} - t_{14} - t_{24}}
\left[g_{h f f}^R
\dfrac{g_{h h h} }{s - m_h^2}
+
g_{H f f}^R
\dfrac{g_{h h H} }{s - m_H^2}
\right]
\nonumber\\
%%%%%%%%%%%%%%%%%%%%%%%%%
&& + 
\dfrac{m_\mu \; 
g_{h H^\pm H^\mp} \cdot 
g_{h f f}^R
\cdot 
g_{h f f}^{L+R} 
}
{\left(t - m_\mu^2\right)
\left(3 m_{H^\pm}^2 + 2 m_\mu^2 
- s_{34} - t_{14} - t_{24}\right)}
\nonumber\\
%%%%%%%%%%%%%%%%%%%%%%%%%
&& +
\dfrac{4 \pi \alpha 
\cdot 
g_{h f f}^R
\cdot 
\left(s_{34} + t - m_h^2 
- m_{H^\pm}^2 - m_\mu^2\right)
}
{\left(t - m_\mu^2\right)
\left(m_h^2 + 3 m_{H^\pm}^2 
+ 2 m_\mu^2 - s_{34} 
- t_{14} - t_{24}\right)}
\nonumber\\
%%%%%%%%%%%%%%%%%%%%%%%%%
&& 
+
\dfrac{g_{H H^\pm H^\mp}}
{m_h^2 - m_H^2 + 3 m_{H^\pm}^2 
+ 2 m_\mu^2 - s_{34} - t_{14} - t_{24}}
\left[
 g_{h f f}^R
\dfrac{ g_{H h h}}{s - m_h^2}
+
g_{H f f}^R
\dfrac{ g_{H H h} }{s - m_H^2}
\right]
\nonumber\\
%%%%%%%%%%%%%%%%%%%%%%%%%
&& + 
\dfrac{m_\mu\; 
g_{H H^\pm H^\mp}
\cdot 
g_{H f f}^R
\cdot 
g_{h f f}^{L+R}
}
{\left(t - m_\mu^2\right)
\left(m_h^2 - m_H^2 + 3 m_{H^\pm}^2 
+ 2 m_\mu^2 - s_{34} 
- t_{14} - t_{24}\right)}
\nonumber\\
%%%%%%%%%%%%%%%%%%%%%%%%%
&& + 
\dfrac{g_{H^\pm H^\mp V}}
{m_h^2 + 3 m_{H^\pm}^2 
+ 2 m_\mu^2 - m_Z^2 
- s_{34} - t_{14} -t_{24}}
\left[
C_{10}
\dfrac{m_\mu }{s - m_Z^2} 
+ C_2\dfrac{1}{t - m_\mu^2}
\right]
\nonumber\\
%%%%%%%%%%%%%%%%%%%%%%
&& + 
\dfrac{g_{h H^\pm H^\mp}}
{2 m_h^2 + m_{H^\pm}^2 
+ 2 m_\mu^2 - s_{34} - t - u}
\left[
g_{h f f}^R
\dfrac{g_{h H^\pm H^\mp}}{s - m_h^2}
+
g_{H f f}^R
\dfrac{g_{H H^\pm H^\mp} }
{s - m_H^2}
\right]
%%%%%%%%%%%%%%%%%%%%%
\nonumber\\
&& + 
\dfrac{m_\mu \; g_{h H^\pm H^\mp} 
\cdot 
g_{H^\pm H^\mp V}
\cdot 
g_{V f f}^{R-L}
}
{\left(s - m_Z^2\right)
\left(2 m_h^2 + m_{H^\pm}^2 
+ 2 m_\mu^2 - s_{34} - t - u\right)}
\nonumber\\
%%%%%%%%%%%%%%%%%%%%%%
&& +
\dfrac{\pi \alpha c_{\beta - \alpha}}
{s_W^2\left[2\left(m_h^2 + m_{H^\pm}^2 
+ m_\mu^2\right) - m_W^2 
- s_{34} - t - u\right]}
\times
\nonumber\\
&&
\hspace{5.7cm}
\times
\left[
\sqrt{2} g_{H^\pm f f}
+
C_4
\dfrac{c_{\beta - \alpha} 
g_{h f f}^R }{s - m_h^2}
-
C_{11}
\dfrac{s_{\beta - \alpha} 
g_{H f f}^R }{s - m_H^2}
\right]
\nonumber\\
%%%%%%%%%%%%%%%%%%%%%%%
&& +
\dfrac{\sqrt{2} \pi \alpha c_{\beta - \alpha} 
\left(s_{34} - m_h^2 - m_{H^\pm}^2\right)
g_{H^\pm f f}
}
{s_W^2 t_{14}
\left[2\left(m_h^2 + m_{H^\pm}^2 + m_\mu^2\right)
- m_W^2 - s_{34} - t - u\right]}
\nonumber\\
%%%%%%%%%%%%%%%%%%%%%%%
&& - 
\dfrac{
\left(s_{34} - m_h^2 - m_{H^\pm}^2\right)
g_{h f f}^R g_{H^\pm f f}^2}
{2 t_{14} \left(u - m_\mu^2\right)}
+
\dfrac{m_\mu \; 
g_{h H^\pm H^\mp} g_{h f f}^R
\cdot 
g_{h f f}^{L+R}
}
{\left(u - m_\mu^2\right)
\left(3 m_{H^\pm}^2 + 2 m_\mu^2 
- s_{34} - t_{14} - t_{24}\right)}
% \left(g_{h f f}^L + g_{h f f}^R\right)
\nonumber\\
%%%%%%%%%%%%%%%%%%%%%%%%
&& -
\dfrac{
4 \pi \alpha
\left(s_{34} - m_h^2 
- m_{H^\pm}^2 - m_\mu^2\right)
g_{h f f}^R
}
{\left(u - m_\mu^2\right)
\left(m_h^2 + 3 m_{H^\pm}^2 + 2 m_\mu^2 
- s_{34} - t_{14} - t_{24}\right)}
\nonumber\\
%%%%%%%%%%%%%%%%%%%%%%%%%
&& + 
\dfrac{m_\mu \; 
g_{H H^\pm H^\mp}
\cdot 
g_{H f f}^R
\cdot
g_{h f f}^{L+R}
}
{\left(u - m_\mu^2\right)
\left(m_h^2 - m_H^2 + 3 m_{H^\pm}^2
+ 2 m_\mu^2 - s_{34} - t_{14} - t_{24}\right)}
% \left(g_{h f f}^L + g_{h f f}^R\right)
\nonumber\\
%%%%%%%%%%%%%%%%%%%%%%%%%
&& 
- C_{14}
\dfrac{g_{H^\pm H^\mp V} }
{\left(u - m_\mu^2\right)
\left(m_h^2 + 3 m_{H^\pm}^2 
+ 2 m_\mu^2 - m_Z^2 - s_{34} 
- t_{14} - t_{24}\right)}
\nonumber\\
%%%%%%%%%%%%%%%%%%%%%%%%%
&& 
- 
\dfrac{4 \pi \alpha\; u \; g_{h f f}^R }
{\left(u - m_\mu^2\right)
\left(m_h^2 + 3 m_{H^\pm}^2 
+ 2 m_\mu^2  - s_{34} 
- t_{14} - t_{24}\right)},
\\
% \end{eqnarray}
%%%%%%%%%%%%%%%%%%%%%%%%%%%%%%%%%%%%%
%%%%%%%%%%%%%%%%%%%%%%%%%%%%%%%%%%%%%
% 	
% \begin{eqnarray}
F_2 &=&
g_{h f f}^L
\dfrac{ g_{H^\pm H^\mp h h}}{s - m_h^2}
+
g_{H f f}^L
\dfrac{ g_{H^\pm H^\mp H h}}{s - m_H^2}
+
\dfrac{g_{h H^\pm H^\mp}}
{s_{34} - m_{H^\pm}^2}
\left[
g_{h f f}^L
\dfrac{g_{h H^\pm H^\mp} }{s - m_h^2}
+
g_{H f f}^L
\dfrac{g_{H H^\pm H^\mp} }{s - m_H^2}
\right]
\nonumber\\
%%%%%%%%%%%%
&& 
+
\dfrac{m_\mu \;
g_{h H^\pm H^\mp} 
g_{H^\pm H^\mp V}
\cdot 
g_{V f f}^{L-R}
}
{\left(s - m_Z^2\right) 
\left(s_{34} - m_{H^\pm}^2\right)}
+
\dfrac{\pi \alpha c_{\beta - \alpha}}
{s_W^2 \left(s_{34} - m_W^2\right)}
\left[
C_8
\cdot 
g_{H f f}^L
\dfrac{s_{\beta - \alpha}  }{s - m_H^2}
-
C_5
\cdot 
g_{h f f}^L 
\dfrac{c_{\beta - \alpha} }{s - m_h^2}
\right]
\nonumber\\
%%%%%%%%%%%%%
&& 
+
\dfrac{g_{h H^\pm H^\mp} }
{3 m_{H^\pm}^2 + 2 m_\mu^2 
- s_{34} - t_{14} - t_{24}}
\left[
 g_{h f f}^L
\dfrac{g_{h h h}}
{s - m_h^2}
+g_{H f f}^L
\dfrac{g_{h h H} }
{s - m_H^2}
\right]
\nonumber\\
%%%%%%%%%%%%%
&&
+ 
\dfrac{m_\mu 
\;
g_{h H^\pm H^\mp} g_{h f f}^R
\cdot
g_{h f f}^{L+R}
}
{\left(t - m_\mu^2\right)
\left(3 m_{H^\pm}^2 + 2 m_\mu^2 
- s_{34} - t_{14} - t_{24}\right)
}
\nonumber\\
%%%%%%%%%%%%%%
&& 
+
\dfrac{4 \pi \alpha 
\left(s_{34} + t - m_h^2 
- m_{H^\pm}^2 - m_\mu^2\right)
g_{h f f}^L}
{\left(t - m_\mu^2\right)
\left(m_h^2 + 3 m_{H^\pm}^2 
+ 2 m_\mu^2 - s_{34} - t_{14} 
- t_{24}\right)}
\nonumber\\
%%%%%%%%%%%%%%
&& +
\dfrac{g_{H H^\pm H^\mp}}
{m_h^2 - m_H^2 + 3 m_{H^\pm}^2 
+ 2 m_\mu^2 - s_{34} - t_{14} - t_{24}}
\left[
g_{h f f}^L
\dfrac{g_{h h H} }{s - m_h^2}
+
g_{H f f}^L
\frac{g_{h H H} }{s - m_H^2}
\right]
\nonumber\\
%%%%%%%%%%%%%%%
&& 
+
\dfrac{
m_\mu \; 
g_{H H^\pm H^\mp} g_{H f f}^L
\cdot 
g_{h f f}^{L+R}
}
{\left(t - m_\mu^2\right)
\left(m_h^2 - m_H^2 + 3 m_{H^\pm}^2 
+ 2 m_\mu^2 - s_{34} 
- t_{14} - t_{24}\right)}
\nonumber\\
%%%%%%%%%%%%%%%
&& 
+
\dfrac{g_{H^\pm H^\mp V}}
{m_h^2 + 3 m_{H^\pm}^2 + 2 m_\mu^2 
- m_Z^2 - s_{34} - t_{14} - t_{24}}
\left[
C_3
\dfrac{1}{t - m_\mu^2}
-
C_{10}
\dfrac{m_\mu }{s - m_Z^2}
\right]
\nonumber\\
%%%%%%%%%%%%%%%
&&
+ 
\dfrac{g_{h H^\pm H^\mp}}
{2 m_h^2 + m_{H^\pm}^2 
+ 2 m_\mu^2 - s_{34} - t - u}
\left[
g_{h f f}^L
\dfrac{g_{h H^\pm H^\mp} }{s - m_h^2}
+
g_{H f f}^L
\dfrac{g_{H H^\pm H^\mp} }{s - m_H^2}
\right]
\nonumber\\
%%%%%%%%%%%%%%
&& +
\dfrac{
m_\mu \; 
g_{h H^\pm H^\mp} g_{H^\pm H^\mp V}
\cdot g_{Vff}^{L-R}
}
{\left(s - m_Z^2\right)
\left(2 m_h^2 + m_{H^\pm}^2 + 2 m_\mu^2 
- s_{34} - t - u\right)}
\nonumber\\
%%%%%%%%%%%%%%
&&
+ 
\dfrac{m_\mu \; 
g_{h H^\pm H^\mp} g_{H^\pm f f}^2}
{t_{14} \left(2 m_h^2 + m_{H^\pm}^2 
+ 2 m_\mu - s_{34} - t - u\right)}
\nonumber\\
%%%%%%%%%%%%%%
&&
+ 
\dfrac{\pi \alpha c_{\beta - \alpha}}
{s_W^2 \left[2\left(m_h^2 + m_{H^\pm}^2
+ m_\mu^2\right) - m_W^2 - s_{34} - t - u\right]}
\times 
\nonumber\\
&& 
\hspace{4.3cm}
\times
\left[
C_4
\frac{c_{\beta - \alpha} g_{h f f}^L }{s - m_h^2}
-
C_{11}
\frac{s_{\beta - \alpha} g_{H f f}^L }{s - m_H^2}
- 
C_6
\frac{\sqrt{2} m_\mu^2 g_{H^\pm f f} }{t_{14}}
\right]
\nonumber\\
%%%%%%%%%%%%%%%%
&& +
\dfrac{m_\mu g_{h H^\pm H^\mp} g_{H^\pm f f}^2}
{\left(s_{34} - m_{H^\pm}^2\right)
\left(m_h^2 + 2 m_{H^\pm}^2 + 3 m_\mu^2 
- s - t_{24} - u\right)}
\nonumber\\
%%%%%%%%%%%%%%%%%%
&& 
- C_{12}
\dfrac{\sqrt{2} \alpha \pi
\; c_{\beta - \alpha} 
\; 
g_{H^\pm f f}}
{s_W^2\; 
\left(s_{34} - m_{H^\pm}^2\right)
\left(m_h^2 + 2 m_{H^\pm}^2 
+ 3 m_\mu^2 - s - t_{24} - u\right)}
\nonumber\\
%%%%%%%%%%%%%%%%
&& +
\dfrac{g_{H^\pm f f}^2 }
{\left(t - m_\mu^2\right)
\left(m_h^2 + 2 m_{H^\pm}^2 
+ 3 m_\mu^2 - s - t_{24} - u\right)}
\left[
m_\mu^2 g_{h f f}^L 
+ 
m_\mu^2 g_{h f f}^R
+
C_{13}
\frac{g_{h f f}^L }{2}
\right]
\nonumber\\
%%%%%%%%%%%%%%%%%%%
&& 
+
\dfrac{
m_\mu^2 
\; g_{H^\pm f f}^2
\cdot g_{h f f}^{L+R}
}
{t_{14} \left(u - m_\mu^2\right)}
+
\dfrac{m_\mu \; 
g_{h H^\pm H^\mp} 
\; g_{h f f}^L
\cdot 
g_{h f f}^{L+R}
} {\left(u - m_\mu^2\right)
\left(3 m_{H^\pm}^2 + 2 m_\mu^2 
- s_{34} - t_{14} 
- t_{24}\right)}
\nonumber\\
%%%%%%%%%%%%%%%%
&& 
-
\dfrac{
4 \pi \alpha
\left(s_{34} - m_h^2 
- m_{H^\pm}^2 - m_\mu^2 \right)
g_{h f f}^L
}
{\left(u - m_\mu^2\right)
\left(m_h^2 + 3 m_{H^\pm}^2 
+ 2 m_\mu^2 - s_{34} 
- t_{14} - t_{24}\right)}
\nonumber\\
%%%%%%%%%%%%%%%%
&& +
\dfrac{m_\mu \; 
g_{H H^\pm H^\mp} g_{H f f}^L
\cdot 
 g_{h f f}^{L+R}
}
{\left(u - m_\mu^2\right)
\left(m_h^2 - m_H^2 
+ 3 m_{H^\pm}^2 
+ 2 m_\mu^2 - s_{34} 
- t_{14} - t_{24}\right)}
% \left( g_{h f f}^L + g_{h f f}^R \right)
\nonumber\\
%%%%%%%%%%%%%%%%
&&
-
C_7
\dfrac{g_{H^\pm H^\mp V} }
{\left(u - m_\mu^2\right)
\left(m_h^2 + 3 m_{H^\pm}^2 
+ 2 m_\mu^2 - m_Z^2 - s_{34} 
- t_{14} - t_{24}\right)}
\nonumber\\
%%%%%%%%%%%%%%%%
&& 
+ 
\dfrac{4 \pi \alpha g_{h f f}^L \; u}
{\left(u - m_\mu^2\right)
\left(m_h^2 + 3 m_{H^\pm}^2 
+ 2 m_\mu^2  - s_{34} 
- t_{14} - t_{24}\right)},
\\
%\end{eqnarray}
%%%%%%%%%%%%%%%%%%%%%%%%%%%%%%%%%%%%%%%
%%%%%%%%%%%%%%%%%%%%%%%%%%%%%%%%%%%%%%%
% \begin{eqnarray}
F_3 &=&
-
\dfrac{2 g_{h H^\pm H^\mp} g_{V f f}^R g_{H^\pm H^\mp V}}
{\left(s - m_Z^2\right)\left(s_{34} - m_{H^\pm}^2\right)}
+
\dfrac{g_{h H^\pm H^\mp} g_{h f f}^L g_{h f f}^R}
{\left(t - m_\mu^2\right)
\left(3 m_{H^\pm}^2 + 2 m_\mu^2 
- s_{34} - t_{14} - t_{24}\right)}
\nonumber\\
&& 
+
\dfrac{8 \pi \alpha g_{h H^\pm H^\mp}
}
{s\left(s_{34} - m_{H^\pm}^2\right)}
+
\dfrac{g_{H H^\pm H^\mp} g_{h f f}^R g_{H f f}^L}
{\left(t - m_\mu^2\right)
\left(m_h^2 - m_H^2 + 3 m_{H^\pm}^2 
+ 2 m_\mu^2 - s_{34} - t_{14} - t_{24}\right)}
\nonumber\\
&& 
+ 
\dfrac{g_{h V V} g_{V f f}^R g_{H^\pm H^\mp V}}
{\left(s - m_Z^2\right)
\left(m_h^2 + 3 m_{H^\pm}^2 + 2 m_\mu^2 
- m_Z^2 - s_{34} - t_{14} - t_{24}\right)}
\nonumber\\
&& 
+ 
\dfrac{m_\mu\;
g_{h f f}^R g_{H^\pm H^\mp V}
\cdot 
g_{V f f}^{L-R}
}
{\left(t - m_\mu^2\right)
\left(m_h^2 + 3 m_{H^\pm}^2 + 2 m_\mu^2 
- m_Z^2 - s_{34} - t_{14} - t_{24}\right)}
\nonumber\\
&& 
+ 
\dfrac{g_{h H^\pm H^\mp} g_{H^\pm f f}^2}
{\left(s_{34} - m_{H^\pm}^2\right)
\left(m_h^2 + 2 m_{H^\pm}^2 
+ 3 m_\mu^2 - s - t_{24} - u\right)}
\nonumber\\
&& 
- 
C_1
\dfrac{\sqrt{2} \pi \alpha m_\mu \; 
c_{\beta - \alpha} g_{H^\pm f f} }
{s_W^2\; \left(s_{34} - m_W^2\right)
\left(m_h^2 + 2 m_{H^\pm}^2 + 3 m_\mu^2 
- s - t_{24} - u\right)}
\nonumber\\
&& + 
\dfrac{m_\mu \; g_{h f f}^R g_{H^\pm f f}^2}
{\left(t - m_\mu^2\right)
\left(m_h^2 + 2 m_{H^\pm}^2 
+ 3 m_\mu^2 - s - t_{24} - u\right)}
\nonumber\\
&& -
\dfrac{g_{h H^\pm H^\mp} 
g_{h f f}^L g_{h f f}^R}
{\left(u - m_\mu^2\right)
\left(3 m_{H^\pm}^2 + 2 m_\mu^2 
- s_{34} - t_{14} - t_{24}\right)}
\nonumber\\
&& -
\dfrac{g_{H H^\pm H^\mp} 
g_{h f f}^L g_{H f f}^R
}
{\left(u - m_\mu^2\right)
\left(m_h^2 - m_H^2 + 3 m_{H^\pm}^2 
+ 2 m_\mu^2 - s_{34} - t_{14} - t_{24}\right)}
\nonumber\\
&& 
+ 
\dfrac{
m_\mu \; 
g_{h f f}^L g_{H^\pm H^\mp V}
\cdot 
g_{V f f}^{L-R}
}
{\left(u - m_\mu^2\right)
\left(m_h^2 + 3 m_{H^\pm}^2 + 2 m_\mu^2 
- m_Z^2 - s_{34} - t_{14} - t_{24}\right)}
,
\\
% %%%%%%%%%%%%%%%%%%%%%%%%%%%%%%%%%%%%%%
% %%%%%%%%%%%%%%%%%%%%%%%%%%%%%%%%%%%%%%
% \begin{eqnarray}
F_4 &=& 
-
\dfrac{2 g_{h H^\pm H^\mp} g_{V f f}^L 
g_{H^\pm H^\mp V}}
{\left(s - m_Z^2\right)\left(s_{34} 
- m_{H^\pm}^2\right)}
+
\frac{8 \pi \alpha m_\mu
\cdot 
g_{h f f}^{L+R}
}
{\left(t - m_\mu^2\right)
\left(m_h^2 + 3 m_{H^\pm}^2 
+ 2 m_\mu^2 - s_{34} - t_{14} 
- t_{24}\right)}
\nonumber\\
&& 
+
\dfrac{8 \pi \alpha g_{h H^\pm H^\mp}}
{s\left(s_{34} - m_{H^\pm}^2\right)}
- 
\dfrac{2 g_{h V V} g_{V f f}^L g_{H^\pm H^\mp V}}
{\left(s - m_Z^2\right)
\left(m_h^2 + 3 m_{H^\pm}^2 + 2 m_\mu^2 
- m_Z^2 - s_{34} - t_{14} - t_{24}\right)}
\nonumber\\
&& 
-
\dfrac{2 m_\mu g_{V f f}^L
g_{H^\pm H^\mp V}
\cdot
g_{h f f}^{L+R} 
}
{\left(t - m_\mu^2\right)
\left(m_h^2 + 3 m_{H^\pm}^2 + 2 m_\mu^2
- m_Z^2 - s_{34} - t_{14} - t_{24}\right)}
\nonumber\\
&& 
+ 
\dfrac{ g_{h H^\pm H^\mp} }
{\left(2 m_h^2 + m_{H^\pm}^2 
+ 2 m_\mu^2 - s_{34} - t - u\right)}
\left[
\dfrac{8 \pi \alpha }
{s }
- 
\dfrac{2 
g_{V f f}^L 
g_{H^\pm H^\mp V}}
{\left(s - m_Z^2\right) }
\right]
\nonumber\\
&& + 
\dfrac{8 \pi \alpha m_\mu
\cdot
g_{h f f}^{L+R}
}
{\left(u - m_\mu^2\right)
\left(m_h^2 + 3 m_{H^\pm}^2 + 2 m_\mu^2
- s_{34} - t_{14} - t_{24}\right)}
\nonumber\\
&& 
- 
\dfrac{2 m_\mu g_{V f f}^L 
g_{H^\pm H^\mp V}
\cdot 
g_{h f f}^{L+R}
}
{\left(u - m_\mu^2\right)
\left(m_h^2 + 3 m_{H^\pm}^2 + 2 m_\mu^2 
- m_Z^2 - s_{34} - t_{14} - t_{24}\right)}
,
\\
% \end{eqnarray}
% %%%%%%%%%%%%%%%%%%%%%%%%%%%%%%%%%%%%%%
% %%%%%%%%%%%%%%%%%%%%%%%%%%%%%%%%%%%%%%
% \begin{eqnarray}
F_5 &=&
-
\dfrac{2 g_{h H^\pm H^\mp} g_{V f f}^L g_{H^\pm H^\mp V}}
{\left(s - m_Z^2\right)\left(s_{34} - m_{H^\pm}^2\right)}
+
\dfrac{g_{h H^\pm H^\mp} g_{h f f}^L g_{h f f}^R}
{\left(t - m_\mu^2\right)
\left(3 m_{H^\pm}^2 + 2 m_\mu^2 
- s_{34} - t_{14} - t_{24}\right)}
\nonumber\\
&& 
+
\dfrac{8 \pi \alpha g_{h H^\pm H^\mp}}
{s\left(s_{34} - m_{H^\pm}^2\right)}
+
\dfrac{g_{H H^\pm H^\mp} g_{h f f}^L g_{H f f}^R}
{\left(t - m_\mu^2\right)
\left(m_h^2 - m_H^2 + 3 m_{H^\pm}^2 
+ 2 m_\mu^2 - s_{34} - t_{14} - t_{24}\right)}
\nonumber\\
&& + 
\dfrac{g_{h V V} g_{V f f}^L g_{H^\pm H^\mp V}}
{\left(s - m_Z^2\right)
\left(m_h^2 + 3 m_{H^\pm}^2 + 2 m_\mu^2 
- m_Z^2 - s_{34} - t_{14} - t_{24}\right)
}
\nonumber\\
&& 
+ 
\dfrac{m_\mu \; 
g_{h f f}^R g_{H^\pm H^\mp V}
\cdot 
g_{V f f}^{R-L}
}
{\left(t - m_\mu^2\right)
\left(m_h^2 + 3 m_{H^\pm}^2 + 2 m_\mu^2 
- m_Z^2 - s_{34} - t_{14} - t_{24}\right)}
\nonumber\\
&&
+ 
\dfrac{2\sqrt{2} \pi \alpha m_\mu \; 
c_{\beta - \alpha} g_{H^\pm f f} }
{s_W^2 t_{14}
\left[2 \left(m_h^2 + m_{H^\pm}^2 
+ m_\mu^2\right) - m_W^2 - s_{34} - t - u\right]}
-
\dfrac{m _\mu \; 
g_{h f f}^R g_{H^\pm f f}^2 }{t_{14} 
\left(u - m_\mu^2\right)}
\nonumber\\
%%%%%%%%%%%%%%%%%%%%%
&& 
-
\dfrac{g_{h H^\pm H^\mp} 
\; 
g_{h f f}^L 
\;
g_{h f f}^R}
{\left(u - m_\mu^2\right)
\left(3 m_{H^\pm}^2 + 2 m_\mu^2 
- s_{34} - t_{14} - t_{24}\right)}
\nonumber\\
&& 
-
\dfrac{g_{H H^\pm H^\mp} 
\;
g_{h f f}^R 
\;
g_{H f f}^L}
{\left(u - m_\mu^2\right)
\left(m_h^2 - m_H^2 + 3 m_{H^\pm}^2 
+ 2 m_\mu^2 - s_{34} 
- t_{14} - t_{24}\right)}
\nonumber\\
&& + 
\dfrac{m_\mu \; g_{h f f}^R 
\; g_{H^\pm H^\mp V}
\cdot 
g_{V f f}^{R-L}
}
{\left(u - m_\mu^2\right)
\left(m_h^2 + 3 m_{H^\pm}^2 + 2 m_\mu^2 
- m_Z^2 - s_{34} - t_{14} - t_{24}\right)},
\\
% % \left( g_{V f f}^R - g_{V f f}^L\right)
% \end{eqnarray}
% %%%%%%%%%%%%%%%%%%%%%%%%%%%%%%%%%%%%%%
% %%%%%%%%%%%%%%%%%%%%%%%%%%%%%%%%%%%%%%
% \begin{eqnarray}
F_6 &=&
-
\dfrac{2 g_{h H^\pm H^\mp} g_{V f f}^R
g_{H^\pm H^\mp V}}
{\left(s - m_Z^2\right)\left(s_{34} - m_{H^\pm}^2\right)}
% \nonumber\\
% &&
+
\dfrac{
8 \pi \alpha m_\mu
\; 
\cdot 
g_{h f f}^{L+R}
}
{\left(t - m_\mu^2\right)
\left(m_h^2 + 3 m_{H^\pm}^2 
+ 2 m_\mu^2 - s_{34} - t_{14} - t_{24}\right)}
% \left(
% g_{h f f}^L + g_{h f f}^R
% \right)
\nonumber\\
&& 
+
\dfrac{8 \pi \alpha g_{h H^\pm H^\mp}}
{s\left(s_{34} - m_{H^\pm}^2\right)}
- 
\dfrac{2 g_{h V V} g_{V f f}^R 
g_{H^\pm H^\mp V}}
{\left(s - m_Z^2\right)
\left(m_h^2 + 3 m_{H^\pm}^2 + 2 m_\mu^2 
- m_Z^2 - s_{34} - t_{14} - t_{24}\right)}
\nonumber\\
&& 
-
\dfrac{
2 m_\mu 
\;
g_{V f f}^R 
g_{H^\pm H^\mp V}
\cdot 
g_{h f f}^{L+R}
}
{\left(t - m_\mu^2\right)
\left(m_h^2 + 3 m_{H^\pm}^2 
+ 2 m_\mu^2 - m_Z^2 - s_{34} 
- t_{14} - t_{24}\right)}
% \left(
% g_{h f f}^L + g_{h f f}^R
% \right)
\nonumber\\
&&
+ 
\dfrac{8 \pi \alpha g_{h H^\pm H^\mp}}
{s \left(2 m_h^2 + m_{H^\pm}^2
+ 2 m_\mu^2 - s_{34} - t - u\right)}
+
\dfrac{g_{h H^\pm H^\mp}
g_{H^\pm f f}^2}
{t_{14} \left(2 m_h^2 + m_{H^\pm}^2 
+ 2 m_\mu^2 - s_{34} - t - u\right)}
\nonumber\\
&& 
- 
\dfrac{2 g_{h H^\pm H^\mp} 
g_{V f f}^R g_{H^\pm H^\mp V}}
{\left(s - m_Z^2\right)
\left(2 m_h^2 + m_{H^\pm}^2 
+ 2 m_\mu^2 - s_{34} - t - u\right)}
\nonumber\\
&& 
\nonumber\\
&& 
-C_6
\dfrac{\sqrt{2} \pi \alpha m_\mu \; 
c_{\beta - \alpha} g_{H^\pm f f} }
{s_W^2 t_{14}
\left[2 \left(m_h^2 + m_{H^\pm}^2 
+ m_\mu^2\right) - m_W^2 - s_{34} 
- t - u\right]} 
\nonumber\\
&& 
+
\dfrac{g_{h H^\pm H^\mp} g_{H^\pm f f}^2}
{\left(s_{34} - m_{H^\pm}^2\right)
\left(m_h^2 + 2 m_{H^\pm}^2
+ 3 m_\mu^2 - s - t_{24} - u\right)} 
\nonumber\\
&&
- 
C_6
\dfrac{\sqrt{2} \pi \alpha m_\mu \; 
c_{\beta - \alpha} g_{H^\pm f f} }
{s_W^2 \left(s_{34} - m_W^2\right)
\left(m_h^2 + 2 m_{H^\pm}^2 + 3 m_\mu^2 
- s - t_{24} - u\right)} 
\nonumber\\
&&
+
\dfrac{m_\mu \; 
g_{H^\pm f f}^2
\cdot 
g_{h f f}^{L+R}
}
{\left(t - m_\mu^2\right)
\left(m_h^2 + 2 m_{H^\pm}^2 + 3 m_\mu^2 
- s - t_{24} - u\right)}
% \left(
% g_{h f f}^L + g_{h f f}^R
% \right)
+
\dfrac{m_\mu 
\;
g_{H^\pm f f}^2
}
{t_{14} \left(u - m_\mu^2\right)}
\cdot 
g_{h f f}^{L+R}
\nonumber\\
&& 
+ 
\dfrac{8 \pi \alpha m_\mu
\; 
\cdot 
g_{h f f}^{L+R}
}
{\left(u - m_\mu^2\right)
\left(m_h^2 + 3 m_{H^\pm}^2 + 2 m_\mu^2 
- s_{34} - t_{14} - t_{24}\right)}
% 		\left(
% 		g_{h f f}^L + g_{h f f}^R
% 		\right)
\nonumber\\
&& 
- 
\dfrac{2 m_\mu g_{V f f}^R 
g_{H^\pm H^\mp V}
\cdot 
g_{h f f}^{L+R}
}
{\left(u - m_\mu^2\right)
\left(m_h^2 + 3 m_{H^\pm}^2 
+ 2 m_\mu^2 - m_Z^2 - s_{34} 
- t_{14} - t_{24}\right)},
% 		\left(
% 		g_{h f f}^L + g_{h f f}^R
% \right),
\\
% \end{eqnarray}
%%%%%%%%%%%%%%%%%%%%%%%%%%%%%%%%%%%%%%%
%%%%%%%%%%%%%%%%%%%%%%%%%%%%%%%%%%%%%%%
% \begin{eqnarray}
F_7 &=&
\dfrac{8 \pi \alpha g_{h f f}^L}
{m_h^2 + 3 m_{H^\pm}^2 + 2 m_\mu^2
- s_{34} - t_{14} - t_{24}}
\left[
\frac{1}{t - m_\mu^2}
+
\frac{1}{u - m_\mu^2}
\right]
\nonumber\\
&&
-
\dfrac{2 g_{h f f}^L 
\; g_{H^\pm H^\mp V}}
{m_h^2 + 3 m_{H^\pm}^2 + 2 m_\mu^2 
- m_Z^2 - s_{34} - t_{14} - t_{24}}
\left[
\frac{g_{V f f}^R}{t - m_\mu^2}
+
\frac{g_{V f f}^L}{u - m_\mu^2}
\right]
\nonumber\\
&& -
\dfrac{2\sqrt{2} \pi \alpha 
c_{\beta - \alpha} g_{H^\pm f f}}
{s_W^2 \left(s_{34} - m_W^2\right)
\left(m_h^2 + 2 m_{H^\pm}^2 
+ 3 m_\mu^2 - s - t_{24} - u\right)}
\nonumber\\
&& +
\frac{g_{h f f}^L g_{H^\pm f f}^2}
{\left(t - m_\mu^2\right)
\left(m_h^2 +2 m_{H^\pm}^2 
+ 3 m_\mu^2 - s - t_{24} - u\right)},
\\
% \end{eqnarray}
% %%%%%%%%%%%%%%%%%%%%%%%%%%%%%%%%%%%%%%
% %%%%%%%%%%%%%%%%%%%%%%%%%%%%%%%%%%%%%%
% \begin{eqnarray}
F_8 &=&
\dfrac{8 \pi \alpha g_{h f f}^R}
{m_h^2 + 3 m_{H^\pm}^2 
+ 2 m_\mu^2 - s_{34} - t_{14} - t_{24}}
\left[
\frac{1}{t - m_\mu^2}
+
\frac{1}{u - m_\mu^2}
\right]
\nonumber\\
%%%%%%%%%%%%%%%%%%%%%%
&& 
-
\dfrac{2 g_{h f f}^R 
g_{H^\pm H^\mp V}}
{m_h^2 + 3 m_{H^\pm}^2 
+ 2 m_\mu^2 - m_Z^2 
- s_{34} - t_{14} - t_{24}}
\left[
\frac{g_{V f f}^L }{t - m_\mu^2}
+
\frac{g_{V f f}^R }{u - m_\mu^2}
\right]
\nonumber\\
%%%%%%%%%%%%%%%%%%%%%%%%
&& -
\dfrac{2\sqrt{2} \pi \alpha 
c_{\beta - \alpha} g_{H^\pm f f}}
{s_W^2 t_{14} 
\left[2 \left(m_h^2 + m_{H^\pm}^2 
+ m_\mu^2\right) - m_W^2 
- s_{34} - t - u\right]}
+
\dfrac{g_{h f f}^R g_{H^\pm f f}^2}
{t_{14} \left(u - m_\mu^2\right)},
\end{eqnarray}
%%%%%%%%%%%%%%%%%%%%%%%%%%%%%%%%%%%%%
%%%%%%%%%%%%%%%%%%%%%%%%%%%%%%%%%%%%%
The operator $P_k$ are chosen as folllow:
%%%%%%%%%%%%%%%%%%%%%%%%%%%%%%%%%%%%%%%%%
$P_1 = P_R$, $P_2 = P_L$, 
$P_3 = P_R \slashed{k_3}$, 
$P_4 = P_L \slashed{k_4}$,
$P_5 = P_L \slashed{k_3}$,
$P_6 = P_R \slashed{k_4}$,
$P_7 = P_L (\slashed{k_3} 
\slashed{k_4} - \slashed{k_4}\slashed{k_3} )1$,
and $P_8 = P_R (\slashed{k_3} 
\slashed{k_4} - \slashed{k_4}\slashed{k_3} )$.
In addition, the coefficient $C_i$ 
are of the form:
\begin{eqnarray}
C_1 &=& \frac{m_h^2 - m_{H^\pm}^2}{m_W^2} - 1, \\
C_2 &=& m_\mu^2 g_{h f f}^{L+R}
% \left(g_{h f f}^L + g_{h f f}^R\right) 
g_{V f f}^{R-L}
% \left(g_{V f f}^R - g_{V f f}^L\right)
+
g_{h f f}^R g_{V f f}^L 
\left(m_h^2 + m_{H^\pm}^2 
+ m_\mu^2 - s_{34} - t\right), \\
C_3 &=&
m_\mu^2 
g_{h f f}^{L+R}
g_{V f f}^{L-R}
% \left(g_{h f f}^L + g_{h f f}^R\right) 
% \left(g_{V f f}^L - g_{V f f}^R\right)
+
g_{h f f}^L g_{V f f}^R 
\left(m_h^2 + m_{H^\pm}^2 
+ m_\mu^2 - s_{34} - t\right), 
\\
%%%%%%%%%%%%%%%%%%%%%%%%%%%%%
C_4 &=& m_h^2 + 3 m_{H^\pm}^2 + 2 m_\mu^2 
+
\dfrac{\left(m_h^2 - m_{H^\pm}^2\right)^2}{m_W^2}
- 2 s_{34} - t_{14} - t_{24}
, \\
C_5 &=& m_h^2 - m_{H^\pm}^2 
-
\frac{\left(m_h^2 - m_{H^\pm}^2\right)^2}{m_W^2}
-
t + t_{14} + t_{24} - u, \\
%%%%%%%%%%%%%%%%%%%%%%%%%%%%%%%%%%%%%%%%%%%%%%%%%%%
C_6 &=&  \frac{m_h^2 - m_{H^\pm}^2}{m_W^2} + 1, \\
%%%%%%%%%%%%%%%%%%%%%%%%%%%%%%%%%%%%%%%%%%%%%%%%%%%
C_7 &=&
m_\mu^2
g_{h f f}^{L+R}
g_{V f f}^{R-L}
% \left(g_{h f f}^L + g_{h f f}^R\right) 
% \left(g_{V f f}^R - g_{V f f}^L\right)
+ 
g_{h f f}^L g_{V f f}^L 
\left(m_h^2 + m_{H^\pm}^2 + m_\mu^2
- s_{34} - u\right), \\
%%%%%%%%%%%%%%%%%%%%%%%%%%%%%%%%%%%%%%%%%%%%%%%%%%%
C_8 &=& m_h^2 - m_{H^\pm}^2 -
\dfrac{\left(m_h^2 - m_{H^\pm}^2\right) 
\left(m_H^2 - m_{H^\pm}^2\right)}{m_W^2}
-
t + t_{14} + t_{24} - u, \\
%%%%%%%%%%%%%%%%%%%%%%%%%%%%%%%%%%%%%%%%%%%%%%%%%%%
C_9 &=& - 2 \left(m_h^2 + m_{H^\pm}^2 
+ m_\mu^2\right) + 2 s_{34} + t + u, \\
%%%%%%%%%%%%%%%%%%%%%%%%%%%%%%%%%%%%%%%%%%%%%%%%%%%%
C_{10} &=& g_{h V V} g_{V f f}^{R-L}
% \left(g_{V f f}^R - g_{V f f}^L\right)
+C_9
\frac{\pi \alpha }{c_W s_W^3}, \\
%%%%%%%%%%%%%%%%%%%%%%%%%%%%%%%%%%%%%%%%%%%%%%%%%%%%
C_{11} &=& m_h^2 + 3 m_{H^\pm}^2 + 2 m_\mu^2 
+
\frac{\left(m_h^2 - m_{H^\pm}^2\right)
\left(m_H^2 - m_{H^\pm}^2\right)}{m_W^2}
-
2 s_{34} - t_{14} - t_{24}, \\
%%%%%%%%%%%%%%%%%%%%%%%%%%%%%%%%%%%%%%%%%%%%%%%%%%%%
C_{12} &=&
\dfrac{m_\mu \left(m_h^2 - m_{H^\pm}^2\right)}{m_W^2}
+
t - t_{14}, \\
%%%%%%%%%%%%%%%%%%%%%%%%%%%%%%%%%%%%%%%%%%%%%%%%%%%%%
C_{13} &=& m_h^2 + m_{H^\pm}^2 - s_{34}
- 2 \left(m_h^2 + m_{H^\pm}^2 + m_\mu^2 - s_{34} - t\right),
\\
%%%%%%%%%%%%%%%%%%%%%%%%%%%%%%%%%%%%%%%%%%%%%%%%%%%%%
C_{14} &=& m_\mu^2 \;
g_{h f f}^{L+R}
g_{V f f}^{L-R}
% \left(g_{h f f}^L + g_{h f f}^R\right) 
% \left(g_{V f f}^L - g_{V f f}^R\right)
+ 
g_{h f f}^R g_{V f f}^R 
\left(m_h^2 + m_{H^\pm}^2 
+ m_\mu^2 - s_{34} - u\right).
%%%%%%%%%%%%%%%%%%%%%%%%%%%%%%%%%%%%%%%%%%%%%%%%%%%%%%
\end{eqnarray} 
Where $g_{h f f}^{L+R}=\left(g_{h f f}^L 
+ g_{h f f}^R\right), \; g_{V f f}^{L-R} 
=\left(g_{V f f}^L - g_{V f f}^R\right)$.
The amplitude for the process 
$\mu^{+}\mu^{-} \to H^{\pm}H^{\mp}H$ 
is obtained by substituting all general couplings 
associated with the external $h$ field above 
with the corresponding couplings of $H$ 
to the relevant particles.
%%%%%%%%%%%%%%%%%%%%%%%%%%%%%%%%%%%%%%%%%%
 %%%%%%%%%%%%%%%%%%%%
%%%%%%%%%%%%%%%%%%%%%%%%%%%%%%%%%%%%%%%%%%
\end{document}